\documentclass[12pt]{article}
\usepackage{amsmath,amsfonts,epsfig,amssymb}

\newcommand{\gammabar}{\overline{\gamma}}
\newcommand{\hagamma}{\widehat{\gamma}}
\newcommand{\mubar}{\overline{\mu}}
\newcommand{\omegabar}{\overline{\omega}}
\newcommand{\sigmabar}{\overline{\sigma}}
\newcommand{\dx}{\mbox{d}x}
\newcommand{\dxi}{\mbox{d}\xi}

\newcommand{\dz}{\mbox{d}z}
\newcommand{\du}{\mbox{d}u}
\newcommand{\dy}{\mbox{d}y}

\newcommand{\dt}{\mbox{d}t}
\newcommand{\dtau}{\mbox{d}\tau}
\newcommand{\dN}{\mbox{d}N}

\newcommand{\picturesAB}[3]{
\centerline{\raise 2mm \hbox{\raise #3 \hbox{(a)}}
\hspace*{-.3in}
\psfig{file=#1,height=#3}
\hspace*{.02in}
\raise 2mm \hbox{\raise #3 \hbox{(b)}}
\hspace*{-.3in}
\psfig{file=#2,height=#3}
}}
\newcommand{\picturesABCm}[4]{
\centerline{\raise #4 \hbox{(a)}\!\!\!\!\!
\psfig{file=#1,height=#4}
\hspace*{.06in}
\raise #4 \hbox{(b)}\!\!\!\!\!\!\!\!\!\!\!\!
\psfig{file=#2,height=#4}
\hspace*{.06in}
\raise #4\hbox{(c)}\!\!\!\!\!\!\!\!\!\!\!\!
\psfig{file=#3,height=#4}
}}

\numberwithin{equation}{section}

\begin{document}

\title{Dynamics of polydisperse irreversible adsorption: \\
a pharmacological example}

\author{Radek Erban\!\,\thanks{University of Oxford, Mathematical Institute, 
24-29 St. Giles', Oxford, OX1 3LB, United Kingdom;
e-mail: {\it erban@maths.ox.ac.uk}. This work was supported 
by the Biotechnology and Biological 
Sciences Research Council.}\,\,\!\,\footnotemark[3]
 \and 
 Jonathan Chapman\!\,\footnotemark[1] 
\and
Kerry D. Fisher\!\,\footnotemark[3]
 \and
Ioannis G. Kevrekidis\!\,\thanks{Princeton University,
Department Of Chemical Engineering, PACM \& Mathematics,
Engineering Quadrangle,
Olden Street, Princeton, NJ 08544, USA.}
\and
Leonard W. Seymour\!\,\thanks{Department of Clinical Pharmacology,
University of Oxford, Radcliffe Infirmary, Woodstock Road,
Oxford, OX2 6HE, United Kingdom.}
}

\maketitle
 
{\small \par \noindent
{\it Abstract:} Many drug delivery systems suffer from undesirable
interactions with the host immune system. 
It has
been experimentally established that covalent attachment 
(irreversible adsorption) of 
suitable macromolecules to the surface of the drug carrier can 
reduce such undesirable interactions. 
A fundamental understanding 
of the adsorption process is still lacking. 
In this paper, the classical random  irreversible adsorption 
model is generalized to capture certain essential processes involved
in pharmacological applications, allowing for 
macromolecules of different sizes,
partial overlapping of the tails of macromolecules,
and the influence of reactions with the solvent
on the adsorption process. 
Working in one dimension, an integro-differential evolution equation 
for the adsorption process is derived and the asymptotic behaviour of 
the surface area covered and the number of molecules attached to the 
surface is studied. 
Finally, equation-free dynamic renormalization tools 
are applied to study the asymptotically self-similar behaviour of the 
adsorption statistics.
\par}

\bigskip

\section{Introduction} 

\label{secintro}

Random sequential adsorption (RSA) is a classical model for various
physical, chemical or biological problems \cite{Evans:1993:RCS}. In the
simplest form, RSA processes can be formulated as sequential addition to
a structure of
objects that cannot overlap, and once inserted, cannot move or leave the
structure \cite{Brilliantov:1996:FFO}.
In this paper, we present a pharmacological example in which application of
the RSA model can provide meaningful qualitative insights. Motivated by 
pharmacological applications, we present 
a slight generalization of the classical RSA model to enable us to study the 
effects of polydispersity and partial overlap of adsorbing macromolecules 
on the surface of a virus. We also study the dependence of the adsorption 
process on interactions (reactions) of the adsorbing macromolecules 
with the solvent.

The paper is organized as follows. In Section \ref{secbiology}, we will
introduce the motivating pharmacological example and our questions of
interest. In Section \ref{secmodel}, we will introduce the generalized
random sequential adsorption (gRSA) model suitable for capturing essential
features of the pharmacological problem from Section \ref{secbiology},
which we formulate in one dimension. The analytical results for
this model are presented in Section \ref{secanalysis}. We derive the
governing integro-differential equation for the evolution of gaps between
polymers, and compute the asymptotical properties
of the quantities of interest, namely the number of macromolecules adsorbed 
and the total area (in one dimension,
length) they cover. In Section \ref{seceqfree}, we apply equation-free
methods to the computational study of the system.
The main idea that underlies this
equation-free computer-assisted analysis is the design and execution of
appropriately-initialized short bursts of stochastic simulations; the
results of these are processed to estimate coarse-grained quantities of
interest - in this case the self-similarly evolving shape of the gap
statistics in the problem.
Finally, in Section \ref{secdiscussion}, we discuss the higher dimensional
case and summarize the connections between the theory and the experimental
data.

\section{Pharmacological background}

\label{secbiology}

Many medical conditions such as cancer, heart disease and heritable
disorders (hemo\-phi\-lia, cystic fibrosis, muscular dystrophy etc) have
faulty, mutant genes as an underlying cause.   Healthy, normal genes can be
readily synthesized in the laboratory but introducing them into diseased
cells remains a challenge.  Many research groups are studying viruses,
such as adenovirus, as a means to introduce normal genes into diseased
cells.  For therapeutic use, the virus' own DNA is usually partially or
completely replaced by the gene of interest.  The most common adenovirus
strain used for this purpose is adenovirus type 5 (Ad5), because it is easy
to manipulate and is non-pathogenic in humans \cite{Vorburger:2002:AGT}.
Ad5 has been used with great success to treat diseases in laboratory animals
but the results have not been replicated in humans.  One of the greatest
problems with using Ad5 in humans is the presence of neutralizing
antibodies.  In addition, the viruses often infect non-target cells,
particularly the liver, causing unwanted toxicity.
Our laboratory is exploring the use of hydrophilic polymers such as
poly[N-(2-hydroxypropyl)methacrylamide] (pHPMA) to coat virus particles and
protect them from neutralising antibodies by steric shielding.  This
technique is very effective at protecting the virus and permitting it
to be retargeted to specific cell types 
\cite{Fisher:2001:PCA,Fisher:2000:VSR,Pouton:1998:KIN}.
The polymer has multiple esters along its length that are used to bind to
the amino side chain of lysine residues on the virus surface.  In a coating
reaction the polymers bind randomly to the virus surface until (a) all of
the lysine residues are occupied or (b) lysine residues are rendered
inaccessible
(obscured) by polymer chains. We know little about the orientation of
polymers on the virus surface or how to optimise the coating reaction
because there are no techniques to visualise the orientation of polymers on
the virus surface.

Representative questions one would like to answer are: 
How many polymer molecules will become attached to the viral surface 
by a given time? How large is the
surface area covered by the polymer coat at that time? In this
paper, the theoretical approach is chosen to address these
questions for simplified models of the adsorption process.
Since the adsorption process is driven by the diffusion of molecules to the
surface of the virus, and since the adsorption is effectively irreversible,
a suitable modification of the classical RSA might be applied to model the
process.
It is important to take into account polydispersity in the polymer size.
Even if we prepare the polymer molecules with a specified target molecular
weight, some relatively small molecules of the polymer will always be
present, and they will diffuse faster than the larger molecules.
A smaller molecule can reach the surface of the virus at a higher rate. We
will consider the adsorption of polymers whose diameters are distributed
according to a probability distribution function $P(z).$ The solution is
assumed to be well-mixed. We also define $p(z)$ as the probability
distribution function of the particles which can reach the surface in a
single time step. An important modelling issue lies in a good choice of
$P(z)$ (e.g. it might be the Gamma distribution) and in a realistic relation
between $P(z)$ and $p(z)$. In Section \ref{secmodel}, we simply specify
$p(z)$ (avoiding the above questions). The long term dynamics of the
polydisperse adsorption  depends on the behaviour of $p(z)$ close to zero.
We will use different distributions $p(z)$ given by (\ref{distrlength}).

Finally, the reactive groups of the polymer molecules can also react with
the solvent before reaching the surface.
This is the case for commonly used biocompatible pHPMA reactive polymers
\cite{Subr:2005:CDC}. If all reactive groups of a polymer are already
hydrolyzed then the polymer cannot covalently bind to the surface of the
virus. Hence, we have to consider that only a fraction $r(t)$ of the
polymers is still reactive at time $t$.
Depending on the form of $r(t)$, different polymer coats may be created.
This issue will be discussed in more detail in Section
\ref{sectimedependent}.

Polymers are long flexible molecules \cite{Doi:1996:IPP}.
The pHPMA polymer molecule has (one or more) reactive
group(s) which can react with the primary amino groups on the viral surface.
As a result, a polymer molecule becomes covalently
(irreversibly) attached to the surface at a point.
The rest of the polymer is not attached (unless another covalent bond is
created) and it freely ``wiggles" in the space above the viral surface.
Having in mind that the ``wiggling tail"
does not perfectly shield the underlying surface, we generalize the
classical RSA model to allow partial overlap of the adsorbing
macromolecules, i.e. we allow some squeezing of the polymers.

Let $N(t)$ be the number of polymers attached to the surface at time $t$.
Let $A(t)$ be the total area of the surface covered by adsorbed polymers at
time $t$.
Since we cover the surface by polymers
of different sizes, there is no obvious relation between $A(t)$ and $N(t)$.
However, both variables $A(t)$ and $N(t)$ are of practical interest as
discussed below.

We cover the surface of the virus by polymers to protect the surface from
unwanted interactions. Hence, the number $A(t)$ gives us the simple
characterization of the area of the surface which is protected by the
polymer coat.
The unwanted interactions are not the only problem which one has to overcome
in order to use viruses as a drug delivery system. Another important task is
to retarget the virus to infect the cells of interest (e.g. cancer cells)
via new receptors. Assuming that we put one ``targeting" group per polymer
molecule, the number of targeting molecules will be equal to $N(t).$

If we considered the adsorption of the same-size nonoverlapping objects 
of the area $a$,
then we would have $A(t)=a N(t)$. In our case, the adsorbing molecules have
different sizes. There is no obvious relation between $A(t)$ and $N(t)$ and
both quantities are of interest. In the following sections, we will present
theoretical approaches to compute the time evolution of $A(t)$ and $N(t)$.

\bigskip

\section{A simple generalization of random sequential adsorption}

\label{secmodel}

Random sequential adsorption has been extensively 
studied during the last several decades \cite{Evans:1993:RCS}.
The theoretical work is more mature in one
dimension with information in higher dimensions mostly
coming from numerical simulations \cite{Krapivsky:1992:KRS}.
If we consider fixed size objects, then RSA usually starts 
from an empty surface and continues until the time
when no further object can be placed, the so-called 
``jamming limit". 
If the objects to be covered have spherical symmetry
\cite{Swendsen:1981:DRS} then the coverage approaches the 
jamming limit as $t^{-1/d}$ where $t$ is time and $d$ is the 
dimension. 
The asymptotic behaviour can be more complicated
for objects of different shape \cite{Evans:1993:RCS}.

As argued in Section \ref{secbiology}, polydispersity
is often present in real systems. 
If we allow adsorbing
particles (in one spatial dimension)
of arbitrarily small length, then the coverage
approaches the full coverage as $t \to \infty.$
Relatively less is known for polydisperse
adsorption. One-dimensional analytical results
can be found in  \cite{Krapivsky:1992:KRS},
where it is assumed that the 
attached polymers prevent binding of other
polymer molecules that would overlap with them.
In reality, the polymer molecules are stretching
during the adsorption process, creating a polymer
brush (for semitelechelic polymers) after sufficiently 
long time \cite{Milner:1988:TGP,Gennes:1980:CPA}. Thus
each  molecule ``covers" a smaller surface area 
at later times. 
Consequently, it is
possible to adsorb more molecules onto the surface.
Here, we take this fact into account and
we modify the random sequential adsorption algorithm
accordingly. 
We state our generalized random sequential
algorithm (gRSA) in one dimension as follows.

\medskip

\leftskip 8mm
\rightskip 8mm

\noindent
{\bf gRSA algorithm:}
{\it
We consider adsorption of small intervals of different 
sizes onto the interval $[0,1]$, the adsorbing domain. 
At each 
time step, the size of a small interval is chosen randomly 
according to the probability distribution function $p(z).$
We select randomly the position of the center $w$ inside the 
adsorbing domain $[0,1]$ and we make an attempt to place the 
small interval of length $z$ there. 
If the center $w$
of the segment to be adsorbed lies inside a segment already placed, the 
adsorption is rejected. 
If the position of the center $w$ is chosen
in the gap $(x_1,x_2)$ between attached polymers, then the new 
polymer segment is adsorbed with probability $\xi(z,w-x_1,x_2-w)$
where $w-x_1$ and $x_2-w$ are distances of the center $w$ of
the polymer from the endpoints of the gap $(x_1,x_2)$. 
}

\leftskip 0mm
\rightskip 0mm

\medskip

\begin{figure}
\picturesAB{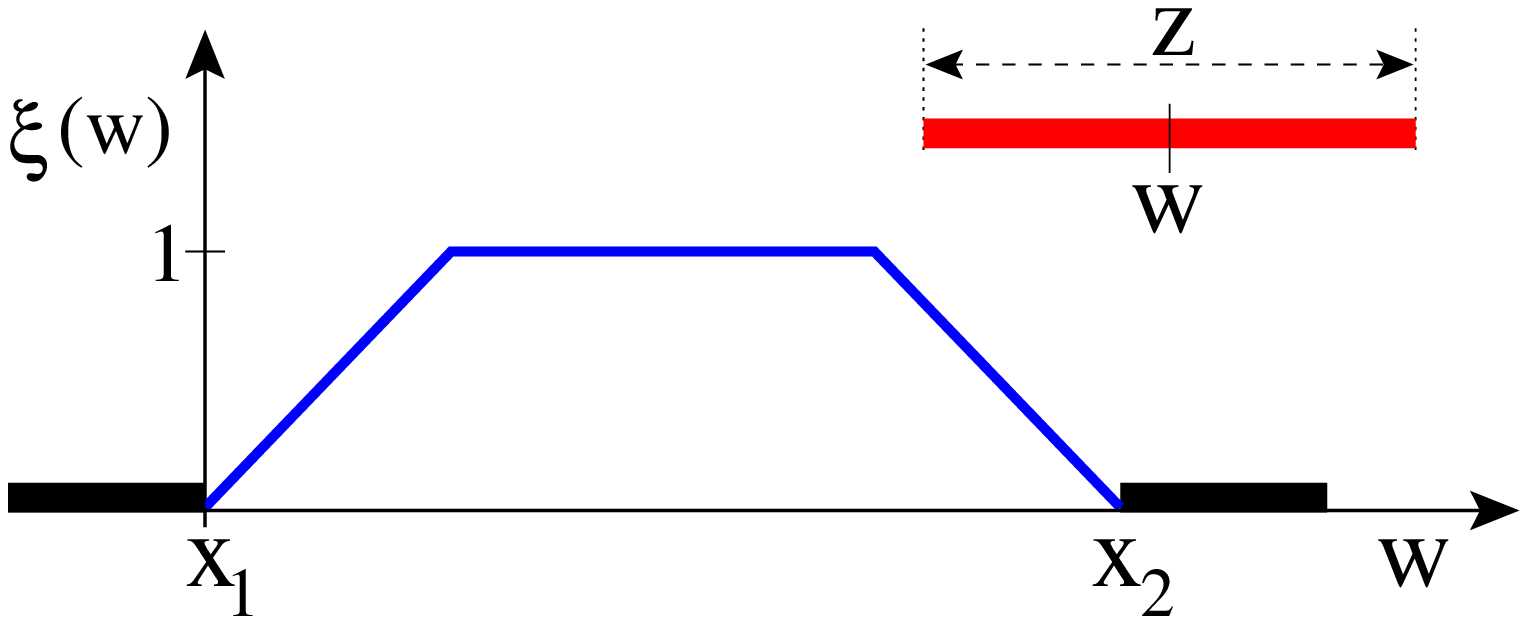}
{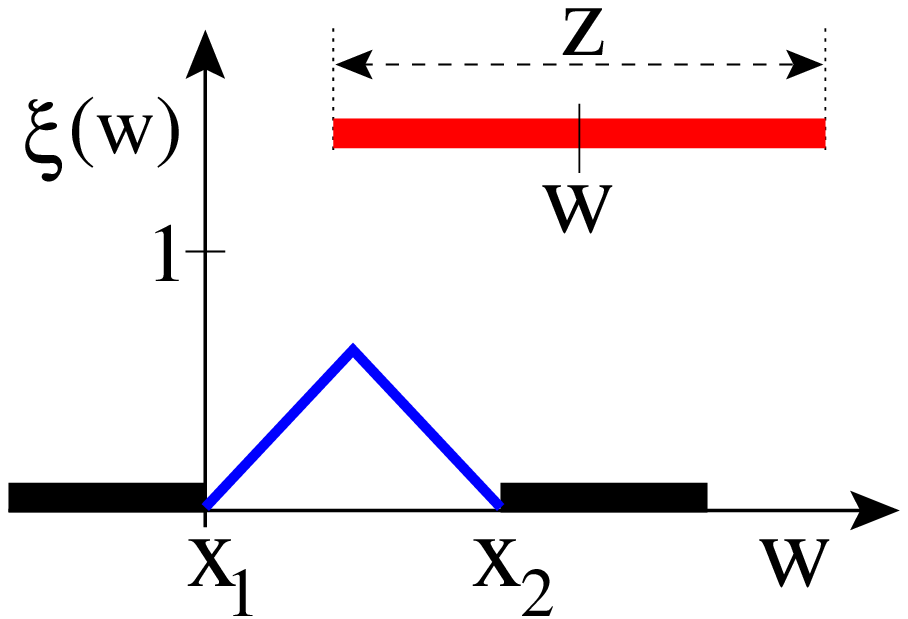}
{1.3in}
\caption{{\it The probability $\xi$ as a function of $w$ for
gRSA model $(\ref{defxi1})$ -- $(\ref{defxi2})$:} 
(a) {\it for the case $z \le x=x_2-x_1$;} 
(b) {\it for the case $z > x=x_2-x_1$}.}
\label{figprobxi}
\end{figure}

\noindent
The parameters of gRSA which have to be specified include the 
probability distribution function $p(z)$ and the
probability $\xi(z,w-x_1,x_2-w)$. 
In what follows,
we assume that the lengths of polymers are distributed according 
to the formula
\begin{equation}
p(z) = 
\left\{
\begin{array}{ll}
(\alpha+1) \varepsilon^{-\alpha-1} z^{\alpha} \quad
& \mbox{for} \; z < \varepsilon, \\
0 & \mbox{for} \; z \ge \varepsilon,
\end{array}
\right. 
\label{distrlength}
\end{equation}
for $\alpha>-1$ and small $\varepsilon \ll 1.$ 
Let us assume that the position of the center $w$ of the 
new polymer is chosen in the gap $[x_1,x_2],$
i.e. $w \in [x_1,x_2]$. 
Then we take the probability (per unit time) of adsorbing the 
polymer segment of length $z \le x = x_1 - x_2$ as
\begin{equation}
\xi(z,w-x_1,x_2-w)
=
\left\{
\begin{array}{ll}
\displaystyle \frac{2 (w - x_1)}{z} \quad & 
\mbox{for} \; w \in \left[ x_1, x_1 + \displaystyle \frac{z}{2} \right]; \\
1 \raisebox{-3.6mm}{\rule{0pt}{10mm}} \quad & 
\mbox{for} \; w \in 
\left[ x_1 + \displaystyle \frac{z}{2}, 
x_2 - \displaystyle \frac{z}{2} \right]; \\
\displaystyle \frac{2 (x_2 - w)}{z} \quad & 
\mbox{for} \; w \in \left[ x_2 - \displaystyle \frac{z}{2}, x_2 \right];
\end{array}
\right.
\label{defxi1}
\end{equation}
and the probability of adsorbing the polymer segment of length 
$z > x$ as
\begin{equation}
\xi(z,w-x_1,x_2-w)
=
\left\{
\begin{array}{ll}
\displaystyle \frac{2 (w - x_1)}{z} \raisebox{-3.6mm}{\rule{0pt}{10mm}} \quad &
\mbox{for} \; w \in \left[ x_1, \displaystyle \frac{x_1 + x_2}{2} \right]; \\
\displaystyle \frac{2 (x_2 - w)}{z} \raisebox{-2.6mm}{\rule{0pt}{10mm}} \quad &
\mbox{for} \; w \in \left[ \displaystyle \frac{x_1 + x_2}{2}, x_2 \right].
\end{array}
\right.
\label{defxi2}
\end{equation}
In the latter case, the maximum probability of adsorption is achieved
for $w=\frac{x_1 + x_2}{2}$, for which 
$\xi(w) = \frac{x}{z}$. 
The formulas (\ref{defxi1}) and (\ref{defxi2}) give
the same probability density function $\xi(\cdot)$ for $z=x$ as is
desirable. 
The plot of $\xi$ as a function of $w$ is given in Figure
\ref{figprobxi}.
Formula (\ref{defxi1}) is shown
in Figure \ref{figprobxi}(a) where the gap size $x=x_2-x_1$
is greater than the length of the new polymer segment $z$.
Formula (\ref{defxi2}) is shown
in Figure \ref{figprobxi}(b) where the gap size $x=x_2-x_1$
is less than the length of the new polymer $z$.

\begin{figure}
\picturesABCm{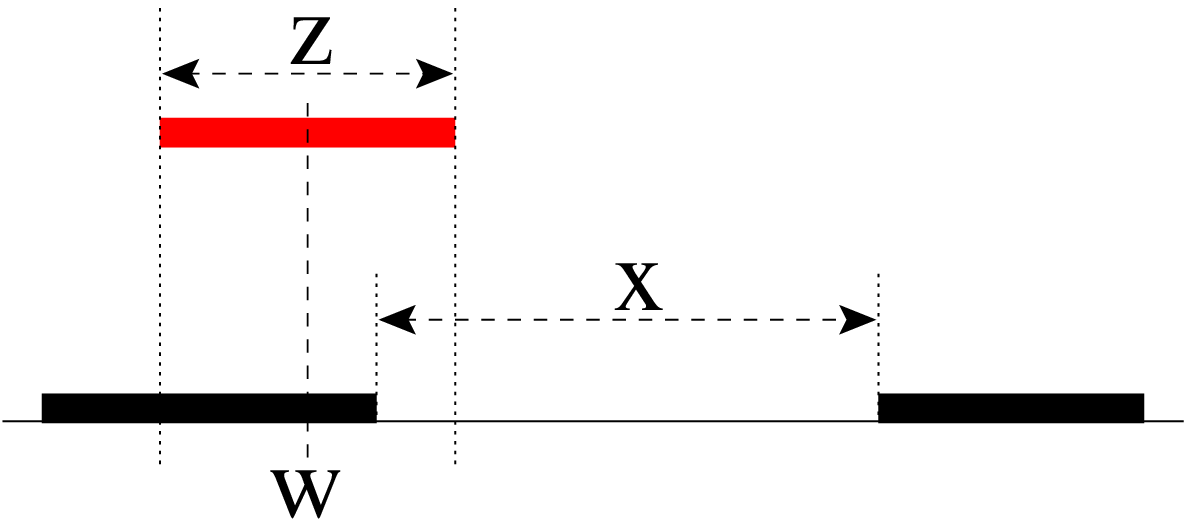}
{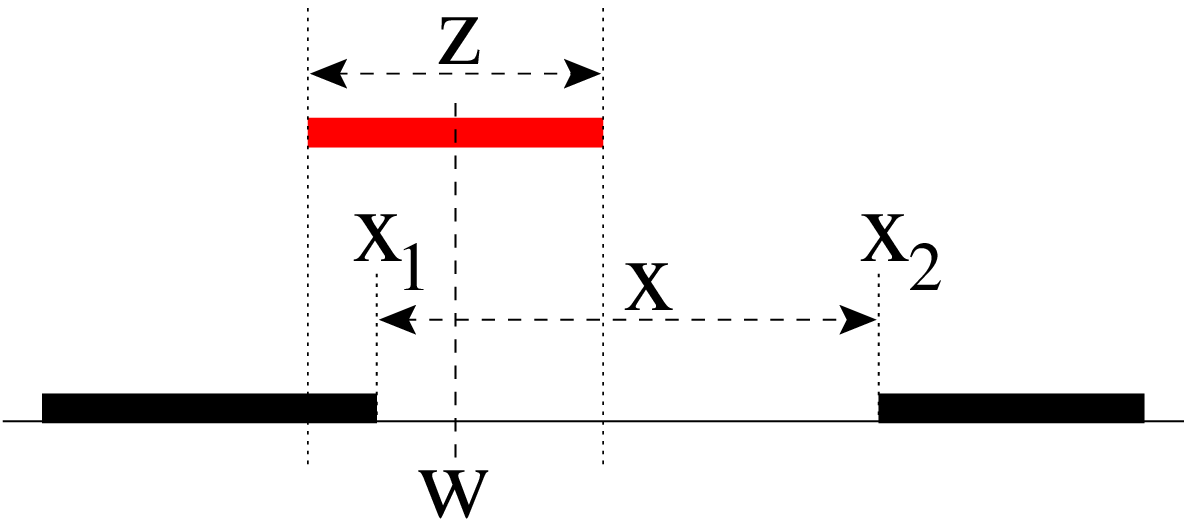}
{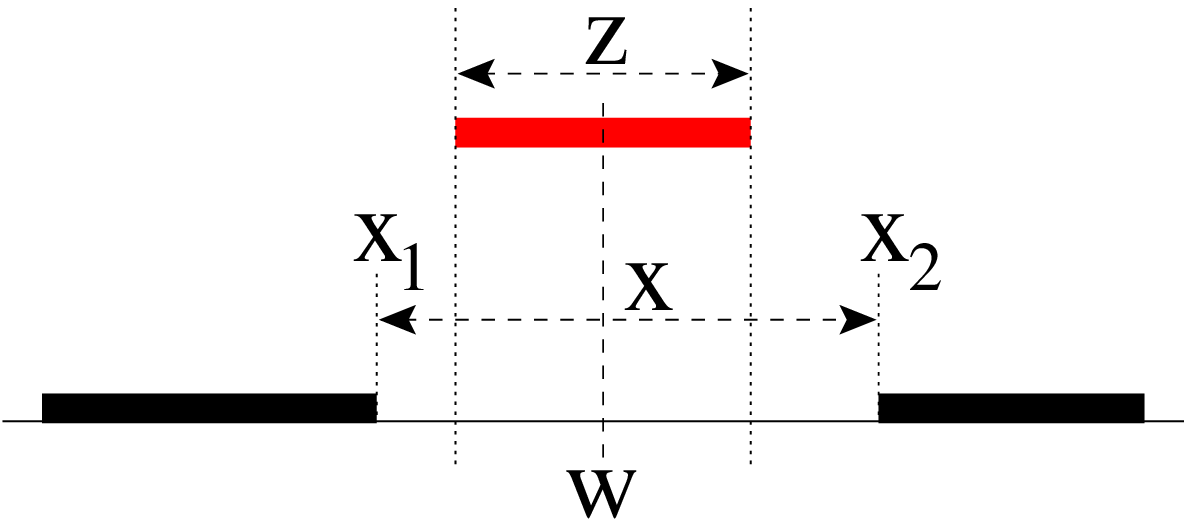}{0.83in}
\caption{{\it Schematic of gRSA.} (a) {\it Polymer is refused;} 
(b) {\it polymer is adsorbed with the
probability $\xi(z,w-x_1,x_2-w)$}; 
(c) {\it polymer is adsorbed.}}
\label{figmodRSA}
\end{figure}

To explain the motivation behind formula (\ref{defxi1}),
three possible cases of the relative position of the new 
(red) interval of the length $z \le x$ and the gap $(x_1,x_2)$ 
are shown in Figure \ref{figmodRSA}.
In Figure \ref{figmodRSA}(a), the red interval is rejected
because its middle point $w$ lies inside a polymer segment which
is already adsorbed to the surface. 
Hence, the probability 
of adsorption is $0$, the same probability as in the 
classical RSA model. 
In Figure \ref{figmodRSA}(c), the red
segment of the length $z$ does not overlap with neighbouring
polymers, and we allow it to be adsorbed with probability $\xi=1$.
Cases in Figure \ref{figmodRSA}(a) and
Figure \ref{figmodRSA}(c) are treated as in the classical RSA 
model.

In Figure \ref{figmodRSA}(b), the center of the red polymer is 
inside the gap but the red polymer overlaps with neighboring 
polymers segments.
This polymer would be rejected by the classical RSA model.
We believe it is more realistic to consider that such a polymer will
be adsorbed with some nonzero probability which continuously
interpolates between the cases shown in Figures \ref{figmodRSA}(a) and
\ref{figmodRSA}(c), i.e. between zero for $w-x_1=0$ and 1 for
$w-x_1=z$. 
Formula (\ref{defxi1}) takes this fact into
account, using simple linear interpolation. 
Formula
(\ref{defxi2}) naturally extends the formula (\ref{defxi1})
for polymer segment lengths greater than the gaps (see also
Figure \ref{figprobxi}).

Having explained the new rules for adsorbing the polymer,
we must also specify what part of the surface is actually
covered. 
We will assume that the new polymer covers only
the intersection of the intervals
\begin{equation}
\left[ w-\frac{z}{2},w+\frac{z}{2} \right]
\bigcap
\left[ x_1, x_2 \right]. 
\label{covered}
\end{equation}
This guarantees that a possibly long, newly
adsorbed polymer will not ``spill over" and cover any
part of the neighboring gaps.

\bigskip

\section{Analysis of gRSA}

\label{secanalysis}

Let $G(x,t)$ be the concentration of gaps (holes) of length 
$x$ at time $t$ and let $C(x,t)$ be the corresponding
cumulative probability
distribution function; that is,
\begin{equation}
C(x,t) = \frac{1}{\int_0^\infty G(y,t) \dy} \int_0^x G(y,t) \dy.
\label{forcdf}
\end{equation}
The total length of the surface that is covered by polymers 
at time $t$, $A(t)$, is directly related to $G(x,t)$ 
by 
\begin{equation}
A(t) 
=
1 - \int_0^1 x G(x,t) \dx.
\label{evolA}
\end{equation}
The number of polymers attached to the surface at time $t$, $N(t)$,
can be also related to $G(x,t)$, as we will see in Section \ref{secevolN}.
Thus, the starting point of the analysis of the system is 
the derivation of the evolution equation for 
 the distribution function of gaps $G(x,t)$.

A gap of length $x$ can be created from a larger gap 
(of length $y>x$) by adsorbing a suitable interval to the 
system. 
Thus the evolution of the concentration of gaps
$G(x,t)$ is given by the equation
\begin{eqnarray}
\frac{\partial G}{\partial t} (x,t)
& = &
-
\; 
G(x,t) 
\int_0^\infty 
\left[ 
\int_0^x \xi(z,u,x-u) 
\du 
\right] 
p(z) 
\dz
\; 
+
\label{evolGgen}
\\
&&
+ 
\int_x^{\infty} 
\left[
\int_0^{2(y-x)} 
2 \,
\xi \left(z, x+\frac{z}{2}, y-x-\frac{z}{2} \right)  
p(z) \dz 
\right]
G(y,t) \dy.
\nonumber
\end{eqnarray}
Using (\ref{defxi1}) and (\ref{defxi2}), equation 
(\ref{evolGgen}) can be rewritten in the following form
$$
\frac{\partial G}{\partial t} (x,t)
=
- G(x,t) 
\int_0^{x} 
\left[ 
\int_0^{z/2} 
\frac{2 u}{z} 
\du 
+
\int_{z/2}^{x-z/2} 
1 \,
\du 
+
\int_{x-z/2}^{x} 
\frac{2 (x-u)}{z} 
\du 
\right] 
p(z) 
\dz 
- 
$$
\begin{equation}
-
G(x,t) 
\int_x^\infty 
\left[ 
\int_0^{x/2} 
\frac{2 u}{z} 
\du 
+
\int_{x/2}^x 
\frac{2 (x-u)}{z} 
\du 
\right] 
p(z) 
\dz 
+ 
\label{evolGgen2}
\end{equation}
$$
+
\int_x^{\infty} 
\int_0^{y-x} 
2 
G(y,t) p(z) 
\dz \dy
+ 
\int_x^{\infty} 
\int_{y-x}^{2(y-x)} 
2 
\frac{2 (y-x) - z}{z} 
G(y,t) p(z) 
\dz \dy.
$$
Hence,
\begin{equation}
\frac{\partial G}{\partial t} (x,t)
=
- G(x,t) 
\int_0^{x} 
\left[
x 
-
\frac{z}{2} 
\right] 
p(z) 
\dz 
- 
G(x,t) 
\int_x^\infty 
\left[ 
\frac{x^2}{2 z} 
\right] 
p(z) 
\dz 
+
\label{evolGgen3}
\end{equation}
$$
+ 
\int_x^{\infty} 
\int_0^{y-x} 
2 
G(y,t) p(z) 
\dz \dy
+ 
\int_x^{\infty} 
\int_{y-x}^{2(y-x)} 
2 
\left[
\frac{2(y-x)}{z} - 1
\right]
G(y,t) p(z) 
\dz \dy.
$$
We assume that the lengths of polymers are
distributed according to formula (\ref{distrlength})
for $\alpha>-1$ and small $\varepsilon \ll 1.$ 
Moreover, we assume 
that there are already 
no holes of the length greater than 
$\varepsilon/2$ in the system, i.e. $G(x,t)=0$ for 
$x>\varepsilon/2.$ 
Then (using (\ref{distrlength})), 
equation (\ref{evolGgen3}) can be rewritten
(for $x < \varepsilon/2$ and $\alpha \ne 0$) as
$$
\frac{\partial G}{\partial t} (x,t)
=
-  
\frac{G(x,t)(\alpha+1)}{\varepsilon^{\alpha+1}} 
\int_0^{x} 
\left[
x 
-
\frac{z}{2} 
\right] 
z^{\alpha}  
\dz 
- 
\frac{G(x,t)(\alpha+1)}{\varepsilon^{\alpha+1}} 
\int_x^{\varepsilon}
\left[ 
\frac{x^2}{2 z} 
\right] 
z^{\alpha}  
\dz 
+
$$
$$
+ 
\frac{2 (\alpha+1)}{\varepsilon^{\alpha+1}}
\int_x^{\infty} 
G(y,t) 
\int_0^{y-x} 
z^{\alpha}  
\dz \dy
+ 
$$
$$
+ 
\frac{2 (\alpha+1)}{\varepsilon^{\alpha+1}}
\int_x^{\infty} 
G(y,t)  
\int_{y-x}^{2(y-x)} 
\left[
\frac{2(y-x)}{z} - 1
\right]
z^{\alpha}  
\dz \dy
$$
which implies
\begin{equation}
\frac{\partial G}{\partial t} (x,t)
=
\frac{x^{\alpha+2} G(x,t)}{\alpha (\alpha+2)\varepsilon^{\alpha+1}}
- 
\frac{x^2 G(x,t)(\alpha+1)}{2 \alpha \varepsilon}
+ 
\frac{2^{\alpha+2}-4}{\alpha \varepsilon^{\alpha+1}} 
\int_x^{\infty}  
G(y,t)
(y-x)^{\alpha+1}
\dy.
\label{evolGgen4}
\end{equation}
If $\alpha=0$, equation (\ref{evolGgen3}) implies
(for $x < \varepsilon/2$)
\begin{equation}
\frac{\partial G}{\partial t} (x,t)
=
- 
\frac{x^2 G(x,t)}{2 \varepsilon}
\left(
\frac{3}{2}
+
\ln \left[ \frac{\varepsilon}{x} \right]
\right)
+ 
\frac{4 \ln 2}{\varepsilon} 
\int_x^{\infty}  
G(y,t)
(y-x)
\dy.
\label{evolGgen5}
\end{equation}
Equation (\ref{evolGgen4}) (or (\ref{evolGgen5}))
is the desired integro-differential equation for $G(x,t)$.
If we solve (\ref{evolGgen4}), we can compute the evolution
of $A(t)$ by (\ref{evolA}). 
The equation for the evolution of $N(t)$ is
given in the next section.

\subsection{Evolution of $N(t)$}

\label{secevolN}

At each time step, an interval of length
between $(z,z+\dz)$ is chosen with probability $p(z)\dz.$
This interval can be placed in any gap of size
$x$ with probability $\int_0^x \xi(z,u,x-u) \du.$
There exist $G(x,t)\dx$ gaps whose size lies
in the interval $(x,x+\dx).$ 
Hence, the integral
$\int_0^\infty [\int_0^x \xi(z,u,x-u) \du]  G(x,t) \dx$ gives the
probability that the randomly chosen position of the 
polymer of length $z$ will be accepted.
Thus the probability of attaching a polymer of any length
at one time step is equal to
\begin{equation}
\int_0^\infty \int_0^\infty 
\left[\int_0^x \xi(z,u,x-u) \du \right] G(x,t) p(z) \dx \dz.
\label{probabilityofattachment}
\end{equation}
Using a continuous approximation for $N(t)$, we find
that $N(t)$ satisfies the following ordinary
differential equation
\begin{equation}
\frac{\dN}{\dt}
=
\int_0^\infty \int_0^\infty 
\left[\int_0^x \xi(z,u,x-u) \du \right] 
 G(x,t) p(z) \dz \dx.
\label{equationforN}
\end{equation}
Taking $p(z)$ to be given by (\ref{distrlength})
and $\xi(z,u,x-u)$ to be given by (\ref{defxi1}) -- (\ref{defxi2}),
and considering the regime where all gaps are
already less than $\varepsilon$ (i.e. $G(x,t) = 0$
for $x > \varepsilon$), we obtain
$$
\int_0^\infty \int_0^\infty 
\left[\int_0^x \xi(z,u,x-u) \du \right] 
 G(x,t) p(z) \dz \dx
= 
$$
$$
\int_0^\infty \int_0^x 
\left[
\int_0^{z/2} 
\frac{2 u}{z}
\du 
+
\int_{z/2}^{x-z/2} 
1 
\du 
+
\int_{x-z/2}^x 
\frac{2 (x -u)}{z}
\du 
\right] 
 G(x,t) p(z) \dz \dx
+
$$
$$
+
\int_0^\infty \int_x^\infty 
\left[
\int_0^{x/2} 
\frac{2 u}{z}
\du 
+
\int_{x/2}^x
\frac{2 (x -u)}{z}
\du 
\right] 
 G(x,t) p(z) 
 \dz \dx
=
$$
$$
=
\int_0^\varepsilon 
 G(x,t) \int_0^x 
\left[
x - \frac{z}{2}
\right] 
p(z) \dz \dx
+
\frac{1}{2} 
\int_0^\varepsilon 
G(x,t) x^2 
\int_x^\infty 
\frac{p(z)}{z} 
\dz 
\dx
=
$$
$$
=
\frac{\alpha+1}{\varepsilon^{\alpha+1}}
\int_0^\varepsilon 
 G(x,t) \int_0^x 
\left[
x - \frac{z}{2}
\right] 
z^{\alpha} \dz \dx
+
\frac{\alpha+1}{2 \varepsilon^{\alpha+1}}
\int_0^\varepsilon 
G(x,t) x^2 
\int_x^\varepsilon
z^{\alpha-1} 
\dz 
\dx
=
$$
$$
=
-
\frac{1}{\alpha(\alpha+2)\varepsilon^{\alpha+1}}
\int_0^\varepsilon 
 G(x,t) x^{\alpha+2}   
\dx
+
\frac{\alpha+1}{2 \alpha \varepsilon}
\int_0^\varepsilon 
G(x,t) x^2
\dx.
$$
Hence
\begin{equation}
\frac{\dN}{\dt}
=
-
\frac{1}{\alpha(\alpha+2)\varepsilon^{\alpha+1}}
\int_0^\varepsilon 
 G(x,t) x^{\alpha+2}   
\dx
+
\frac{\alpha+1}{2 \alpha \varepsilon}
\int_0^\varepsilon 
G(x,t) x^2
\dx.
\label{equationforN2}
\end{equation}
Before analyzing (\ref{evolGgen4}) and (\ref{equationforN2})
further, we summarize some results from the literature
on classical RSA. 

\bigskip

\subsection{Some results for the classical RSA}

\label{classicalRSA}

If we consider particles of the same length
$\varepsilon$ so that  $p(z) = \delta (z - \varepsilon)$,
and if we choose $\xi(z,w-x_1,x_2-w)$ of the form
\begin{equation}
\xi(z,w-x_1,x_2-w)
=
\left\{
\begin{array}{ll}
1 \raisebox{-3.6mm}{\rule{0pt}{8mm}} \quad & 
\mbox{for} \; \;
x_1 + \displaystyle \frac{z}{2} \le w \le 
x_2 - \displaystyle \frac{z}{2}; \\
0 \raisebox{-2.6mm}{\rule{0pt}{8mm}} \quad &
\mbox{otherwise};
\end{array}
\right.
\label{defxi3}
\end{equation}
then our gRSA algorithm reduces to the classical RSA algorithm.
The evolution equation 
(\ref{evolGgen}) can be used to verify known one-dimensional results 
about fixed segment size, non-overlapping random sequential 
adsorption \cite{Krapivsky:1992:KRS},
namely one can show that the jamming limit
is asymptotically approached as $t^{-1}$ \cite{Swendsen:1981:DRS}.

\begin{figure}
\picturesAB{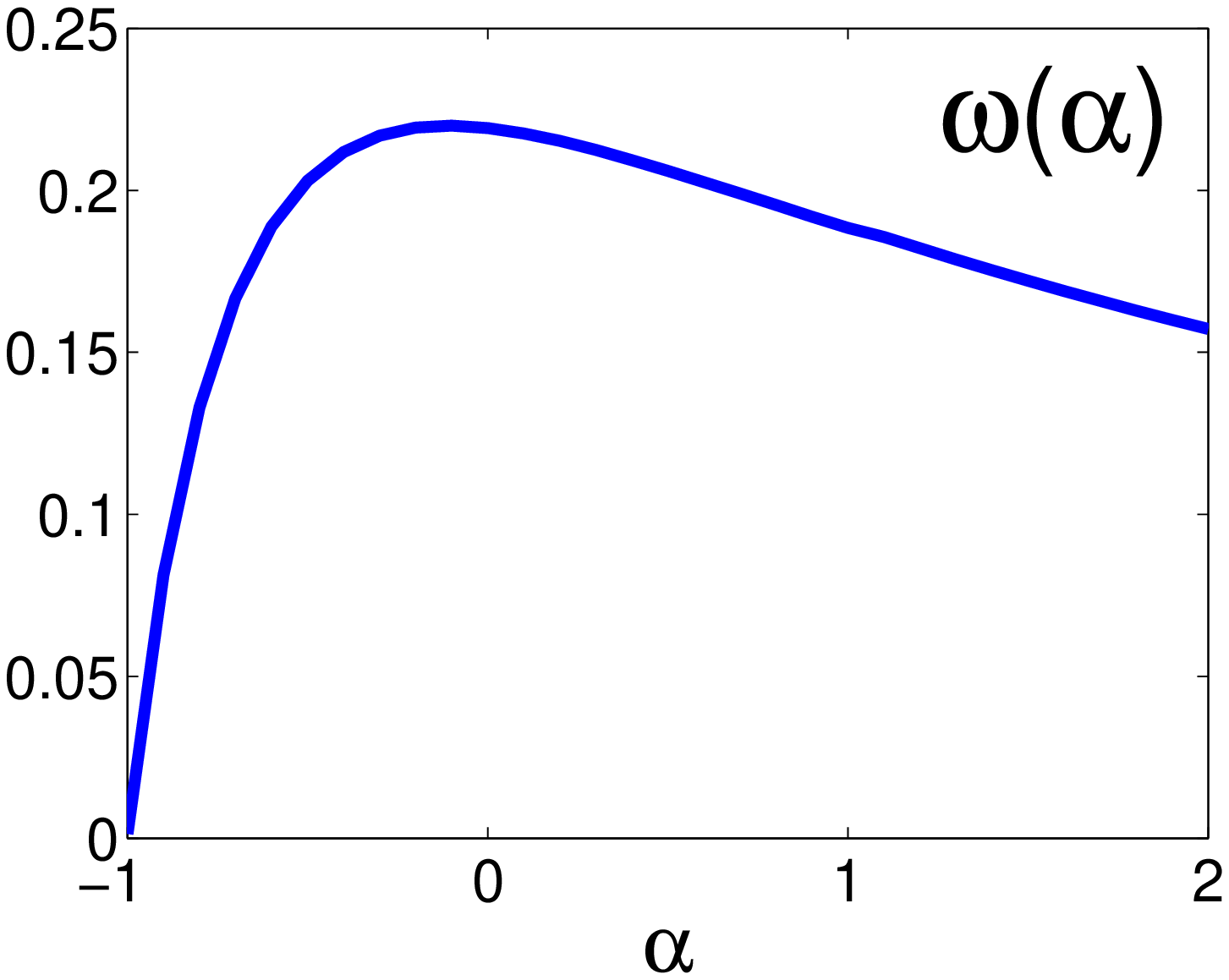}{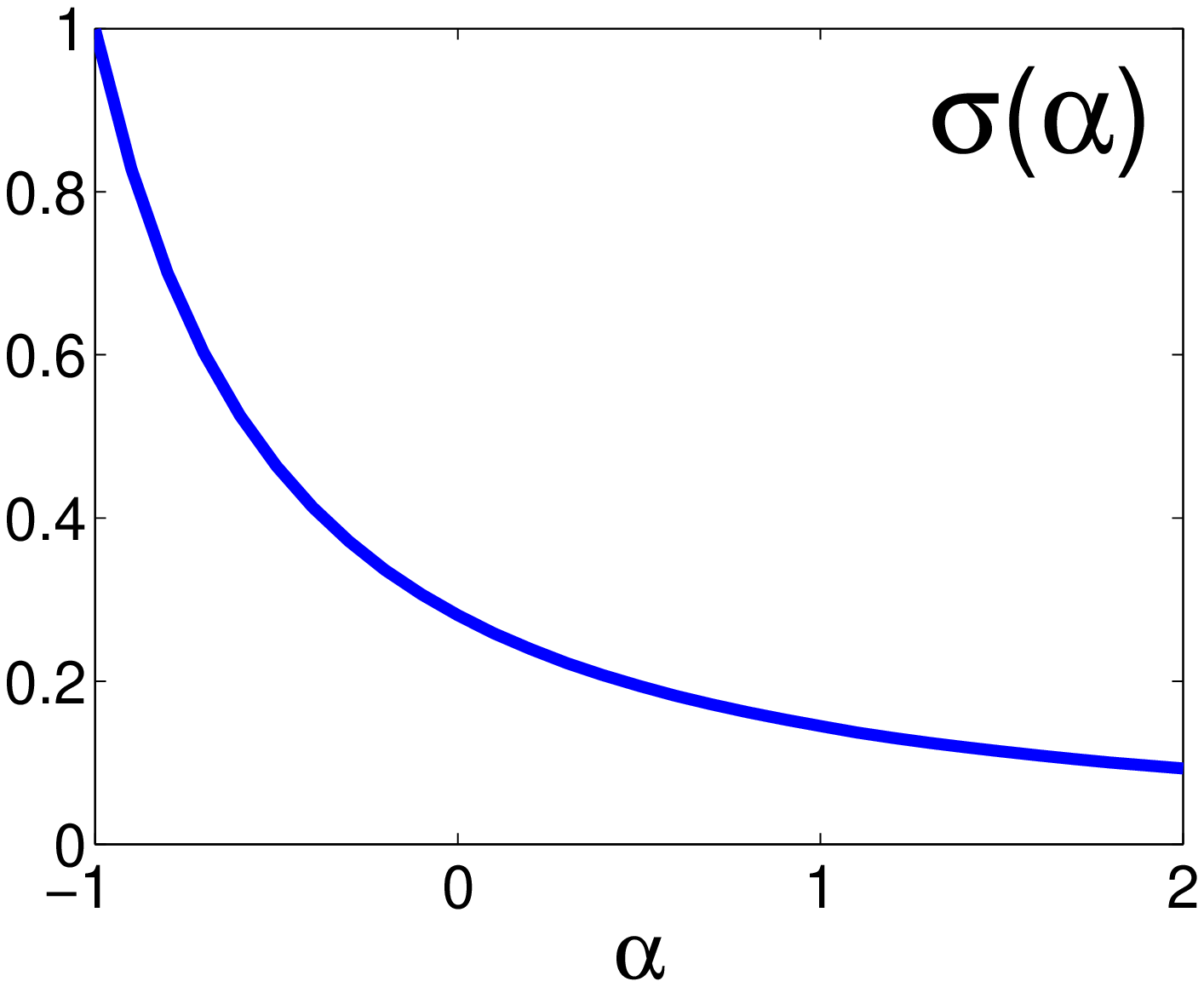}{2.2in}
\caption{(a) {\it
The graph of the exponent $\omega(\alpha)$ given
by $(\ref{scalingAt})$.}
(b) {\it
The graph of the exponent $\sigma(\alpha)$ given
by $(\ref{scalingNt})$.}
}
\label{figgraphomegasigma}
\end{figure}

Random sequential adsorption with a probability 
distribution $p(z)$ given by (\ref{distrlength}) and 
probability $\xi(z,w-x_1,x_2-w)$ given by (\ref{defxi3})
has been studied in \cite{Krapivsky:1992:KRS}.
Then equation (\ref{evolGgen}) for
$x<\varepsilon$ reads as follows (assuming that initially there exist no
holes of length greater than $\varepsilon$
in the system, i.e. $G(x,t)=0$ for $x>\varepsilon$)
\begin{equation}
\frac{\partial G}{\partial t} (x,t)
=
- 
\frac{x^{\alpha+2} G(x,t)}{(\alpha+2)\varepsilon^{\alpha+1}}
+
\frac{2}{\varepsilon^{\alpha+1}} 
\int_x^{x+\varepsilon} G(y,t) (y-x)^{\alpha+1} \dy.
\label{evolGn2}
\end{equation}
The scaling ansatz \cite{Krapivsky:1992:KRS}
for the concentration $G(x,t)$ can be written as
\begin{equation}
G(x,t) \sim t^{a} \Phi \left( x \, t^b \, \right)
\quad
\mbox{for}
\quad
x \ll 1,
\quad
t \gg 1,
\quad
\mbox{and}
\quad
x t^b \; \mbox{finite}.
\label{scalingansatz}
\end{equation}
Defining the moments
\begin{equation}
M_\gamma (t) = \int_0^\infty x^{\gamma} G(x,t) \dx,
\qquad
m_\gamma = \int_0^\infty \xi^{\gamma} \Phi(\xi) \dxi
\label{defmoments}
\end{equation}
and using (\ref{scalingansatz}), we obtain
\begin{equation}
M_\gamma (t) \sim  t^{a - b - b \gamma} m_\gamma. 
\label{scalmoments}
\end{equation}
Moreover, multiplying equation (\ref{evolGn2}) by $x^\gamma$
and integrating over $x$, one can derive
the equation for moments,
\begin{equation}
\frac{\partial M_\gamma}{\partial t} (x,t)
=
\frac{F(\gamma,\alpha)}{\varepsilon^{\alpha+1}} 
M_{\gamma + \alpha + 2}
\quad
\mbox{where}
\quad
F(\gamma,\alpha) 
=
2
B(\gamma+1,\alpha+2)
- 
\frac{1}{\alpha+2},
\label{evolGn3}
\end{equation}
where $B(\cdot,\cdot)$ is Beta function.
We define the function $\hagamma(\alpha)$ implicitly
by the equation $F(\hagamma,\alpha) = 0$. 
If $\gamma$ is equal
to $\hagamma(\alpha)$, then the moment $M_\gamma$
is independent of time. 
Hence, using (\ref{scalmoments}),
we obtain the relation 
$a = b + b \, \hagamma(\alpha)$
between the coefficients of the scaling ansatz (\ref{scalingansatz}) 
and  the parameter $\alpha$ of the model.
Finally, substituting the scaling ansatz (\ref{scalingansatz})
in (\ref{evolGn2}), we find that
$b = (\alpha+2)^{-1}.$ 
Thus, the scaling of moments
(\ref{scalmoments}) can be rewritten in the form 
\begin{equation}
M_\beta (t) \sim  
t^{\mu}
\qquad
\mbox{where}
\quad
\mu
=
\frac{\hagamma(\alpha) - \beta}{\alpha+2}. 
\label{scalmomentsrevised}
\end{equation}
Using (\ref{evolA}) and (\ref{scalmomentsrevised}), we obtain
\begin{equation}
1 - A(t) 
=
\int_0^1 x G(x,t) \dx 
\sim
t^{-\omega(\alpha)}
\qquad
\mbox{where}
\quad
\omega(\alpha)
=
\frac{1-\hagamma(\alpha)}{\alpha+2}. 
\label{scalingAt}
\end{equation}
The graph of the function $\omega(\alpha)$
is given in Figure \ref{figgraphomegasigma}(a).
The equation (\ref{equationforN}) for $p(z)$ given by 
(\ref{distrlength}) and probability $\xi(z,w-x_1,x_2-w)$ 
given by (\ref{defxi3}) reads as follows:
\begin{equation}
\frac{\dN}{\dt}
=
\frac{\varepsilon^{-\alpha-1}}{\alpha+2}
\int_0^\varepsilon 
x^{\alpha+2}
G(x,t) \dx.
\label{equationforN3}
\end{equation}
Using (\ref{scalmomentsrevised}), we obtain (for $\sigma(\alpha)>0$)
\begin{equation}
N(t)
\sim  
t^{\sigma(\alpha)}
\qquad
\mbox{where}
\; \;
\sigma(\alpha)
=
\frac{\hagamma(\alpha)}{\alpha+2}. 
\label{scalingNt}
\end{equation}
The graph of the function $\sigma(\alpha)$
is given in Figure \ref{figgraphomegasigma}(b).

\subsection{Evolution of gRSA}

\label{secalphaminus}

\begin{figure}
\centerline{{\Large $\alpha = - 0.1$}
\qquad \qquad \qquad \qquad \qquad \qquad {\Large $\alpha = - 0.3$}}
\centerline{
\psfig{file=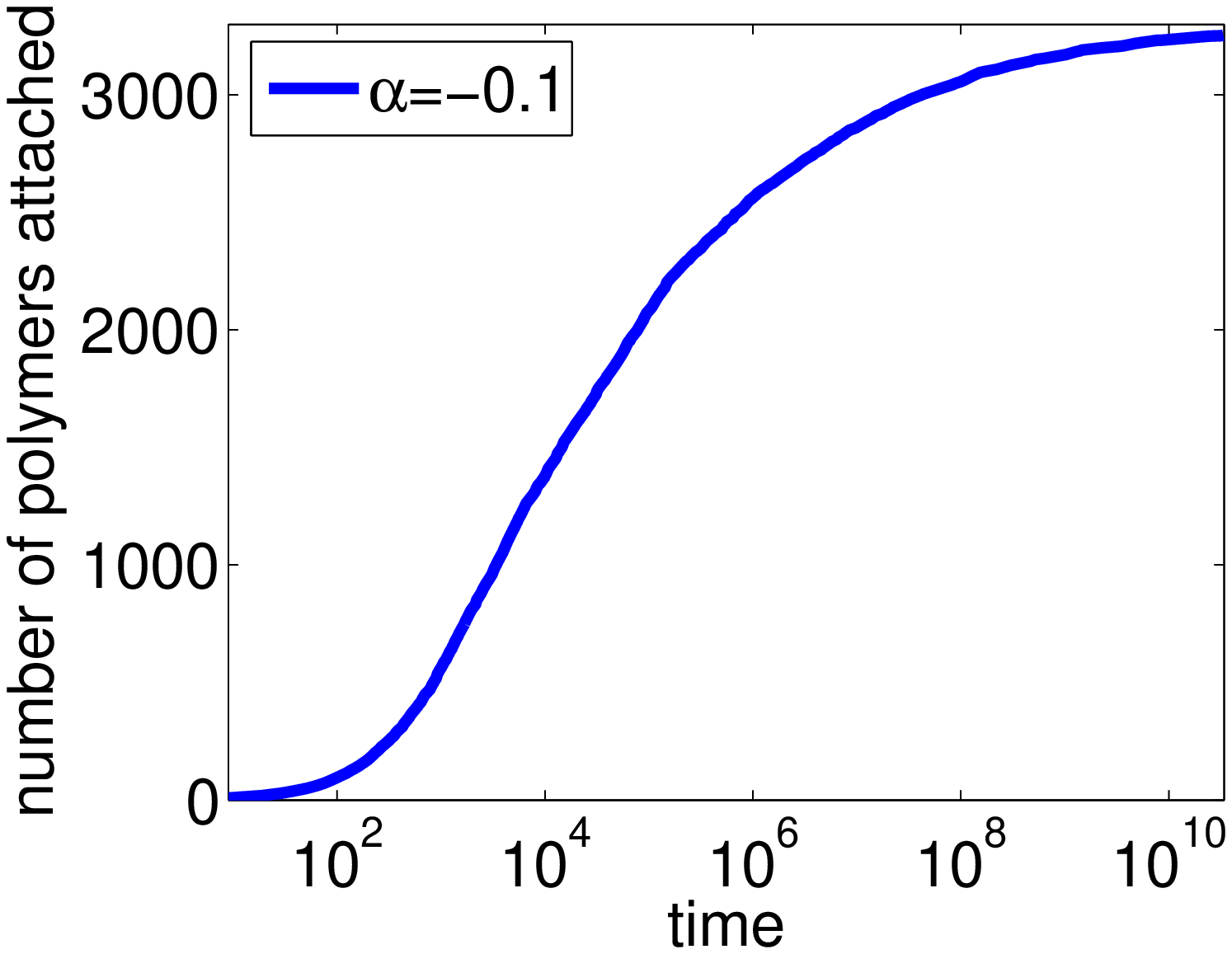,height=2.2in}
\quad
\psfig{file=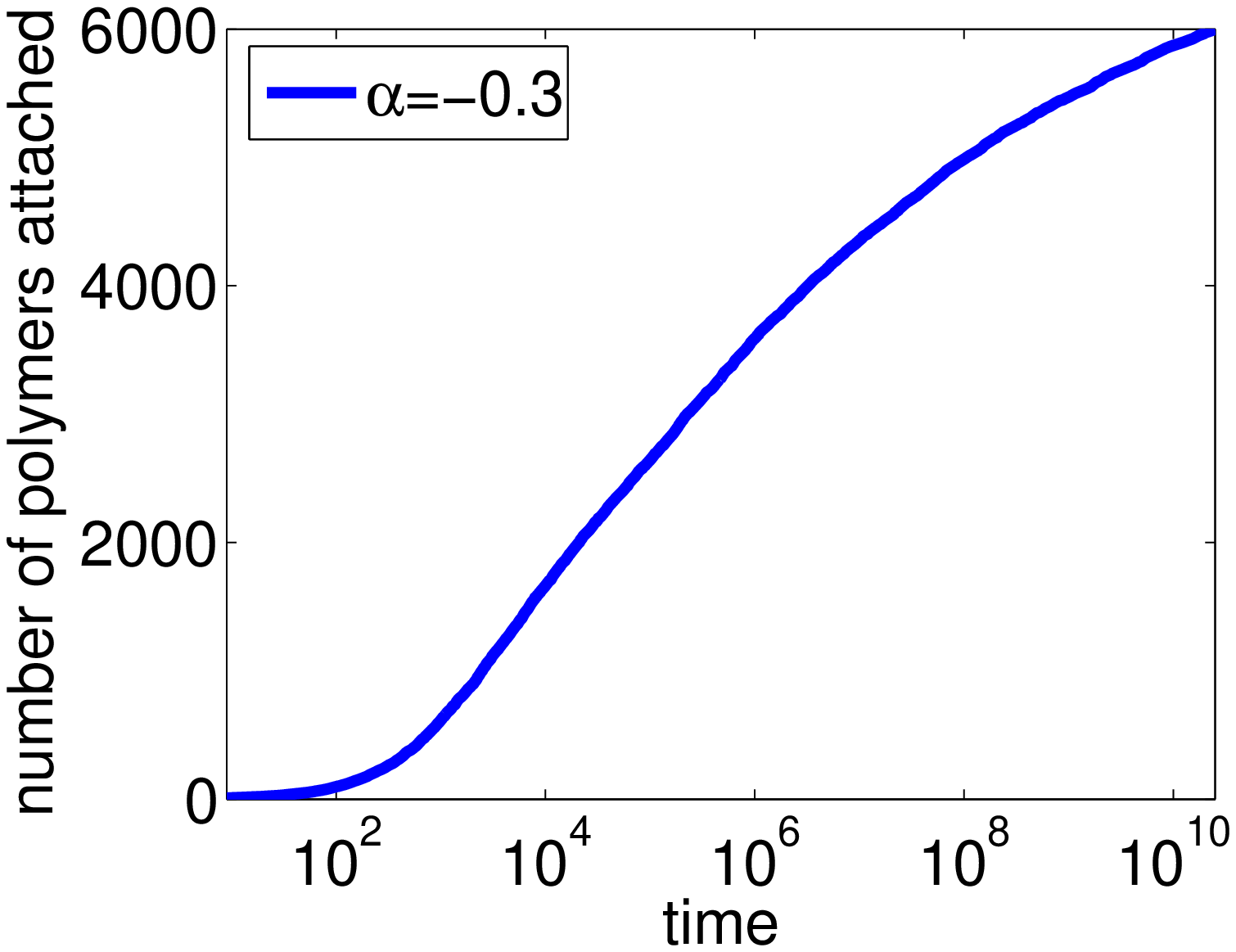,height=2.2in}
}
\smallskip
\centerline{
\psfig{file=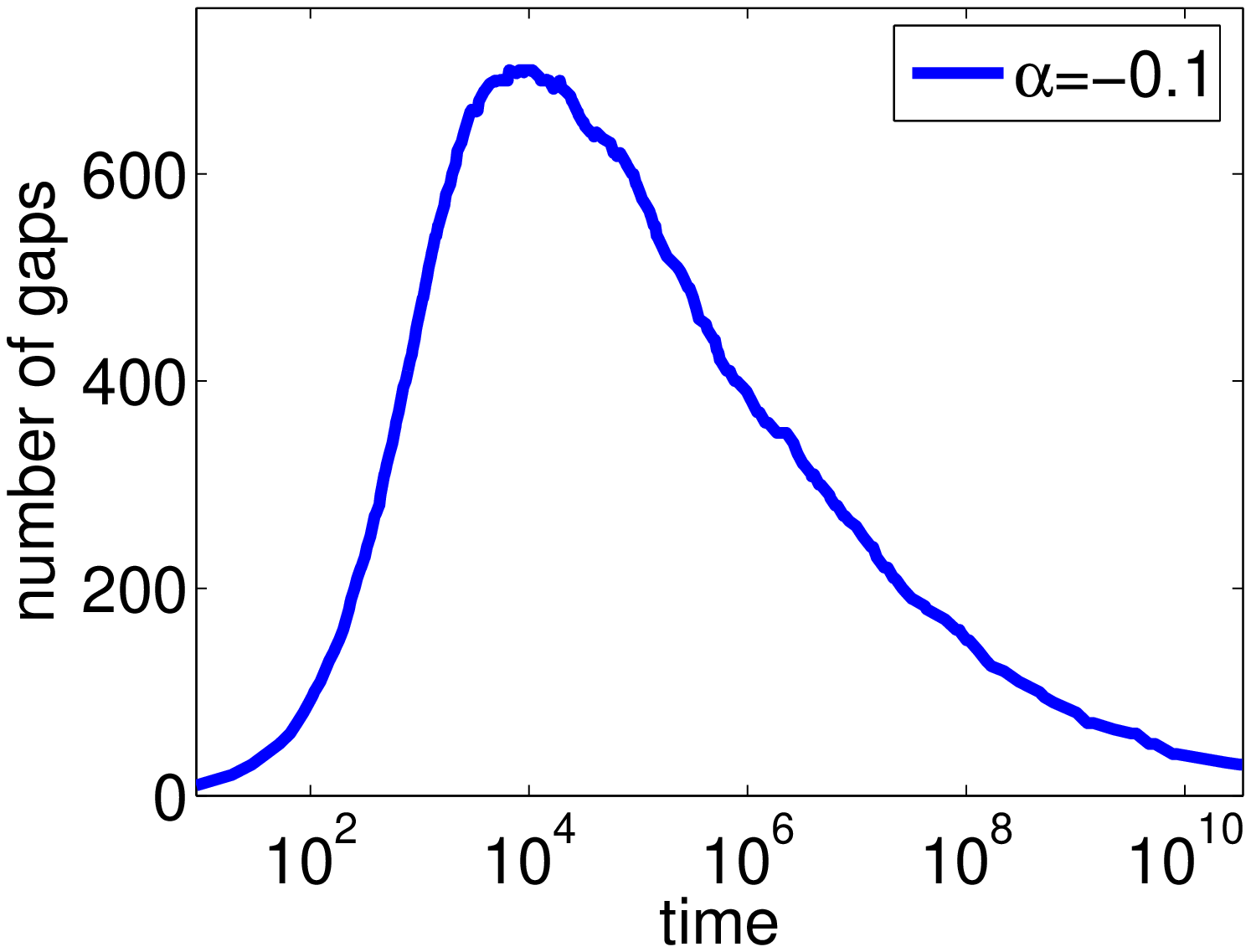,height=2.2in}
\quad
\psfig{file=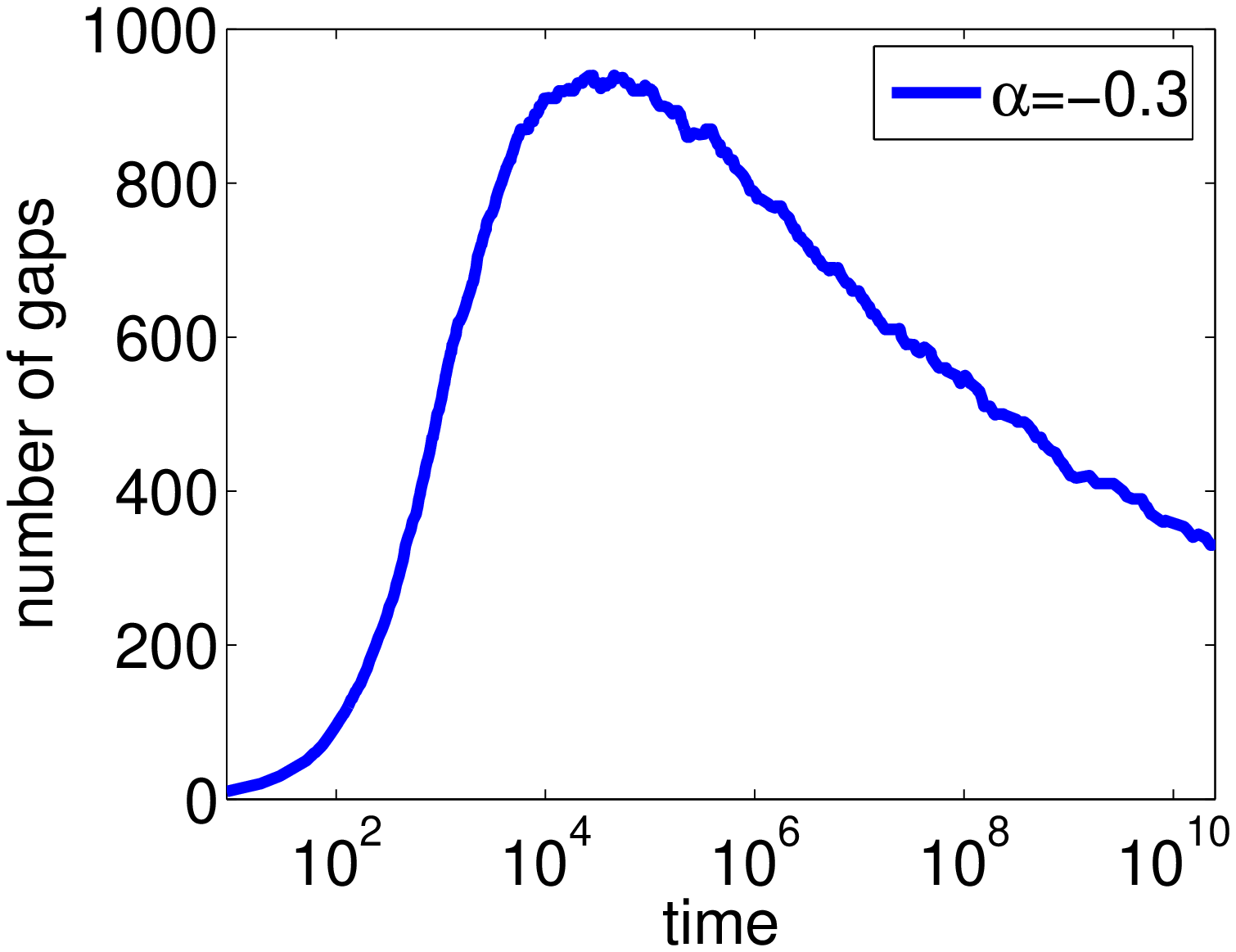,height=2.2in}
}
\smallskip
\centerline{
\psfig{file=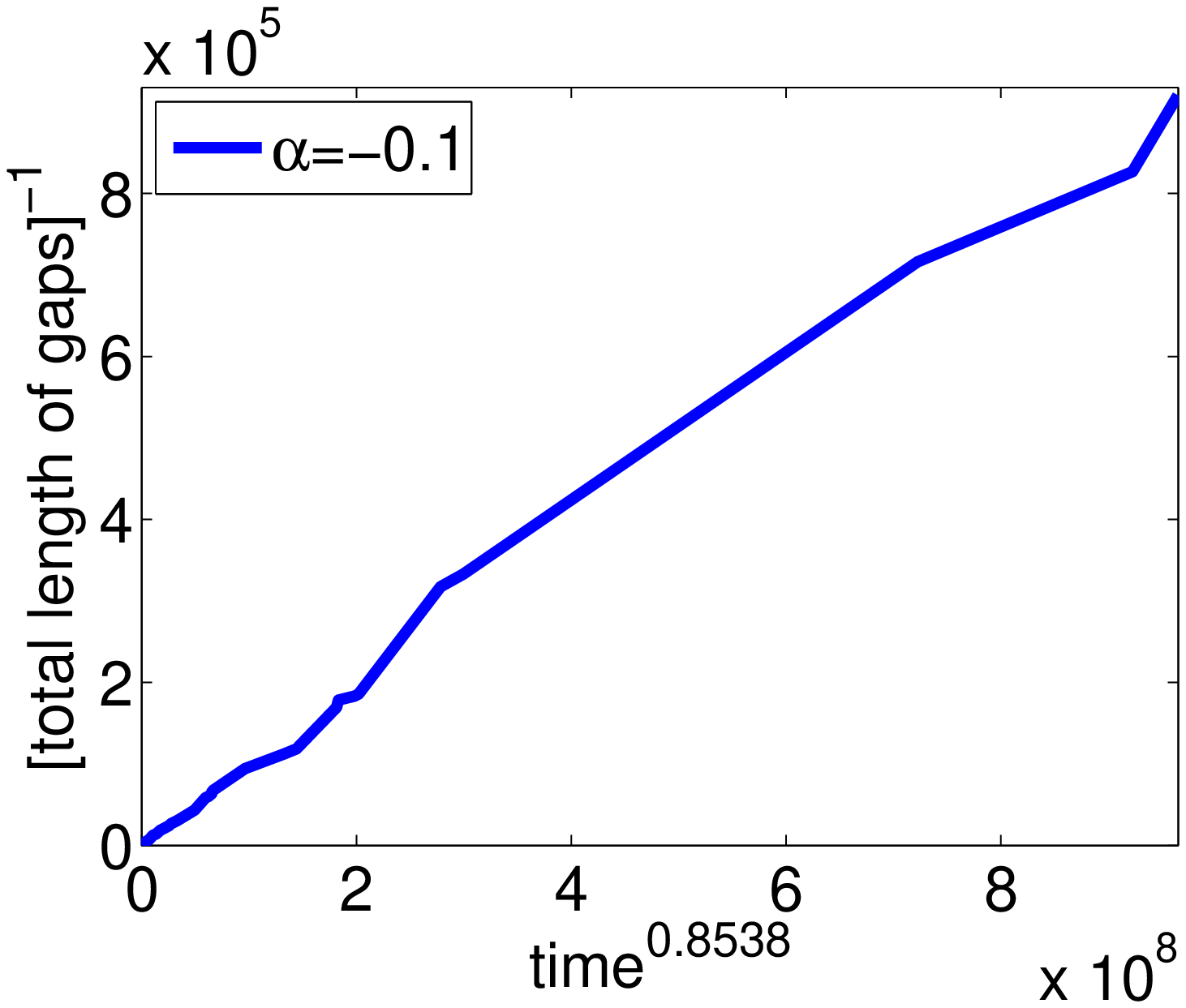,height=2.2in}
\quad
\psfig{file=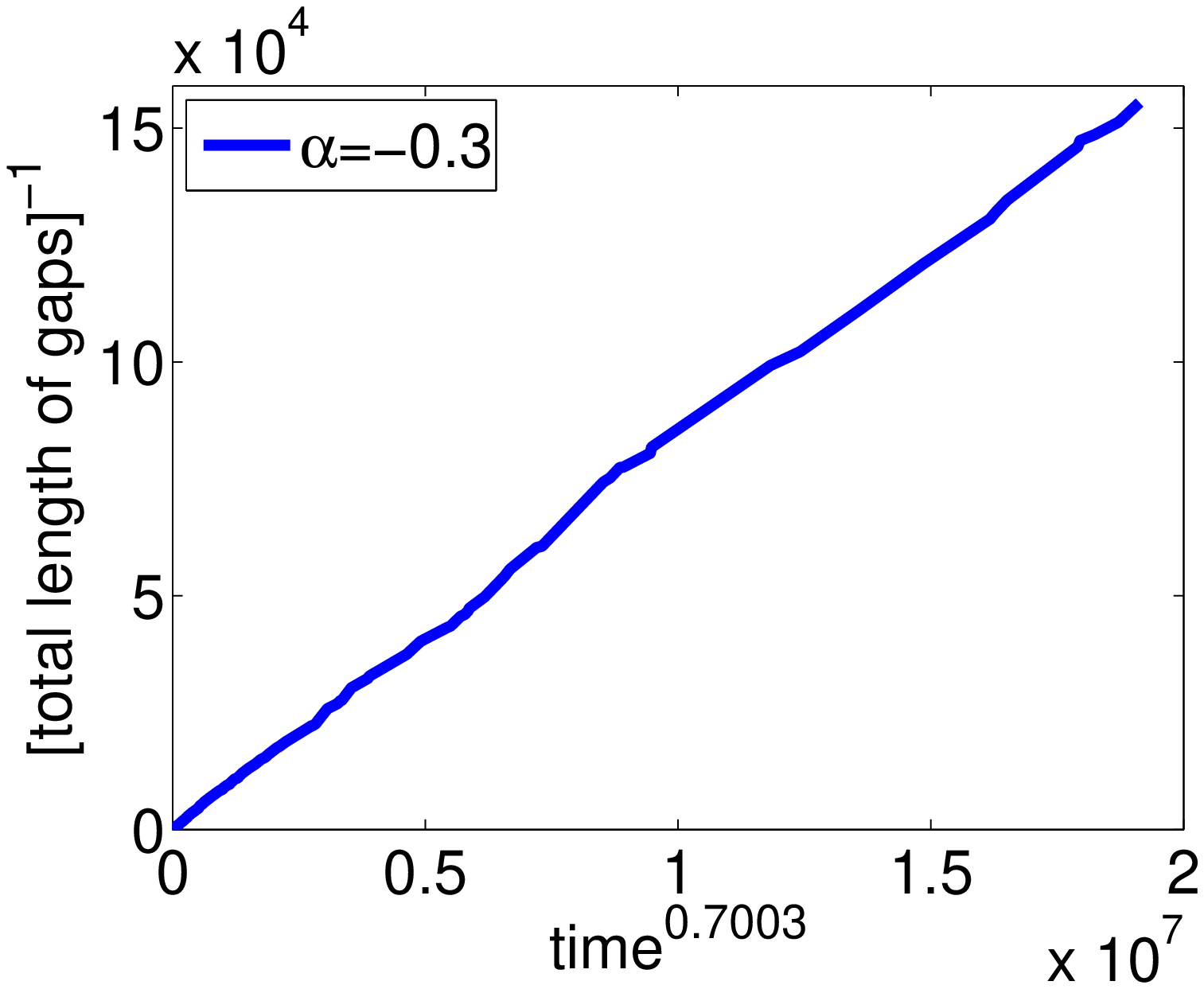,height=2.2in}
}
\caption{{\it gRSA model for $\alpha=-0.1$ (panels on
the left) and $\alpha=-0.3$ (panels on the right).
We plot the time evolution of the number of adsorbed
polymers (top panels) and the time evolution of the number
of gaps (middle panels). The time axis of the top and middle panels
is logarithmic. We also plot the time 
evolution of the quantity $[1-A(t)]^{-1}$ (bottom panels)
where time is scaled according to $(\ref{scalingAtgenRSA})$.
}}
\label{figalphasmall}
\end{figure}

The temporal evolution of the gRSA model is more complex than the cases
discussed in Section \ref{classicalRSA}. 
To see this, we use the moments $M_\gamma(t)$ defined by (\ref{defmoments}).
Multiplying equation (\ref{evolGgen4}) by $x^\gamma$
and integrating over $x$, we can derive
the equation for moments (for $\alpha \ne 0$),
$$
\frac{\partial M_\gamma}{\partial t} (x,t)
= 
\frac{1}{\alpha (\alpha+2)\varepsilon^{\alpha+1}}
M_{\gamma + \alpha + 2}
- 
\frac{\alpha+1}{2 \alpha \varepsilon}
M_{\gamma + 2}
\;
+ 
$$
$$
+
\;
\int_0^\infty
\frac{2^{\alpha+2}-4}{\alpha \varepsilon^{\alpha+1}} 
\int_x^{\infty}  
G(y,t)
x^{\gamma} (y-x)^{\alpha+1}
\dy \dx
=
$$
$$
=
\frac{1}{\alpha (\alpha+2)\varepsilon^{\alpha+1}}
M_{\gamma + \alpha + 2}
- 
\frac{\alpha+1}{2 \alpha \varepsilon}
M_{\gamma + 2}
+ 
\int_0^{\infty}  
\frac{2^{\alpha+2}-4}{\alpha \varepsilon^{\alpha+1}} 
G(y,t)
\int_0^{y}  
x^{\gamma} (y-x)^{\alpha+1}
\dx \dy
=
$$
$$
=
\frac{M_{\gamma + \alpha + 2}}{\alpha (\alpha+2)\varepsilon^{\alpha+1}}
+ 
\frac{2^{\alpha+2}-4}{\alpha \varepsilon^{\alpha+1}} 
\int_0^\infty
G(y,t) y^{\gamma+\alpha+2} \dy
\int_0^{1} \xi^{\gamma} (1-\xi)^{\alpha+1} \dxi
- 
\frac{\alpha+1}{2 \alpha \varepsilon}
M_{\gamma + 2}
=
$$
$$
=
\frac{1}{\varepsilon^{\alpha+1}}
\left(
\frac{2^{\alpha+2}-4}{\alpha} 
B(\gamma+1,\alpha+2)
+ 
\frac{1}{\alpha (\alpha+2)}
\right)
M_{\gamma + \alpha + 2}
- 
\frac{\alpha+1}{2 \alpha \varepsilon}
M_{\gamma + 2}
=
$$
\begin{equation}
=
\frac{1}{\varepsilon^{\alpha+1}}
H(\gamma,\alpha) 
M_{\gamma + \alpha + 2}
- 
\frac{\alpha+1}{2 \alpha \varepsilon}
M_{\gamma + 2},
\label{evolmom1}
\end{equation}
where $B(\cdot,\cdot)$ is Beta function and $H(\gamma,\alpha)$ is 
defined as
\begin{equation}
H(\gamma,\alpha) 
=
\frac{2^{\alpha+2}-4}{\alpha} 
B(\gamma+1,\alpha+2)
+ 
\frac{1}{\alpha (\alpha+2)}.
\label{defHgammaalpha}
\end{equation}
First, consider the case $\alpha<0$; the dominant term on the right-hand 
side of (\ref{evolmom1}) is the term $\varepsilon^{-\alpha-1}
H(\gamma,\alpha) M_{\gamma + \alpha + 2}$. 
The zeroth order moment,
$$
M_0(t) = \int_0^\infty G(x,t) \dx,
$$
gives the total number of gaps at time $t$. 
At leading order, we have 
$$
\frac{\partial M_0}{\partial t} (x,t)
=
\varepsilon^{-\alpha-1}
H(0,\alpha) M_{\alpha + 2}.
$$
There is a constant $\overline{\alpha} \doteq -0.415$ such that
$H(0,\alpha)$ is positive for $\alpha<\overline{\alpha}$
and negative for $\alpha>\overline{\alpha}$. 
We immediately see that 
\begin{equation}
\int_0^\infty G(x,t) \dx \quad \to \quad 0
\qquad \qquad
\mbox{for} \quad \alpha>\overline{\alpha}.
\label{decaygaps}
\end{equation}
To illustrate the result (\ref{decaygaps}), we will execute two
stochastic simulations with the gRSA algorithm. 
We will use (\ref{distrlength}),
(\ref{defxi1}) and (\ref{defxi2})
where $\alpha=-0.1$ or $\alpha=-0.3$. 
We choose $\varepsilon=10^{-3}$. 
The results are given
in Figure \ref{figalphasmall}, where
the time evolution of the number of gaps and
the number of adsorbed polymers are shown. 
Note that we use a
logarithmic scale on the time axis because the long-term dynamics
are very slow. 
For $\alpha=-0.1$, the stochastic
simulation was stopped when $99.9999\%$ of the
surface was covered.
For $\alpha=-0.3$, the stochastic
simulation was stopped when $99.9994\%$ of the
surface was covered. 

\begin{figure}
\picturesAB{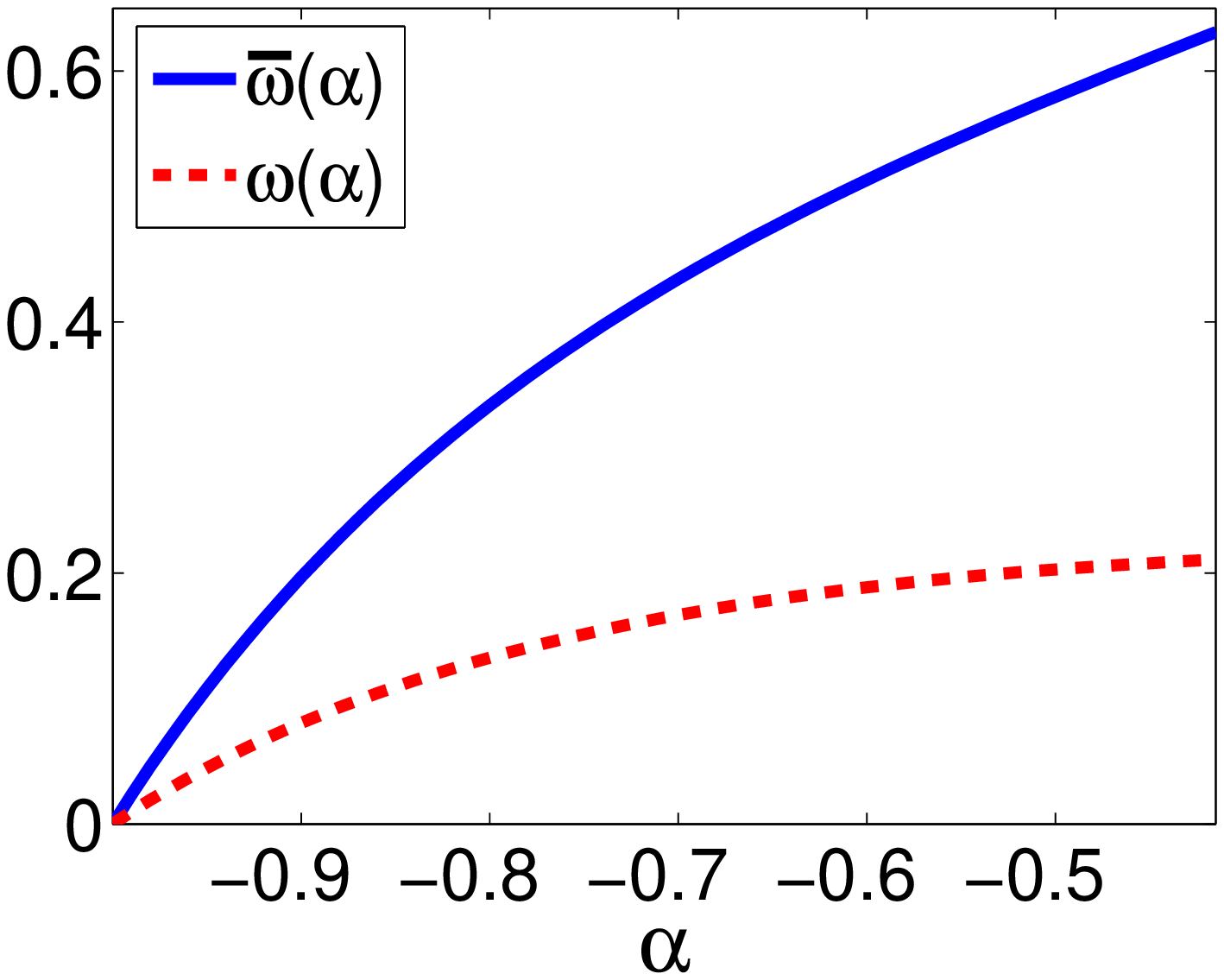}{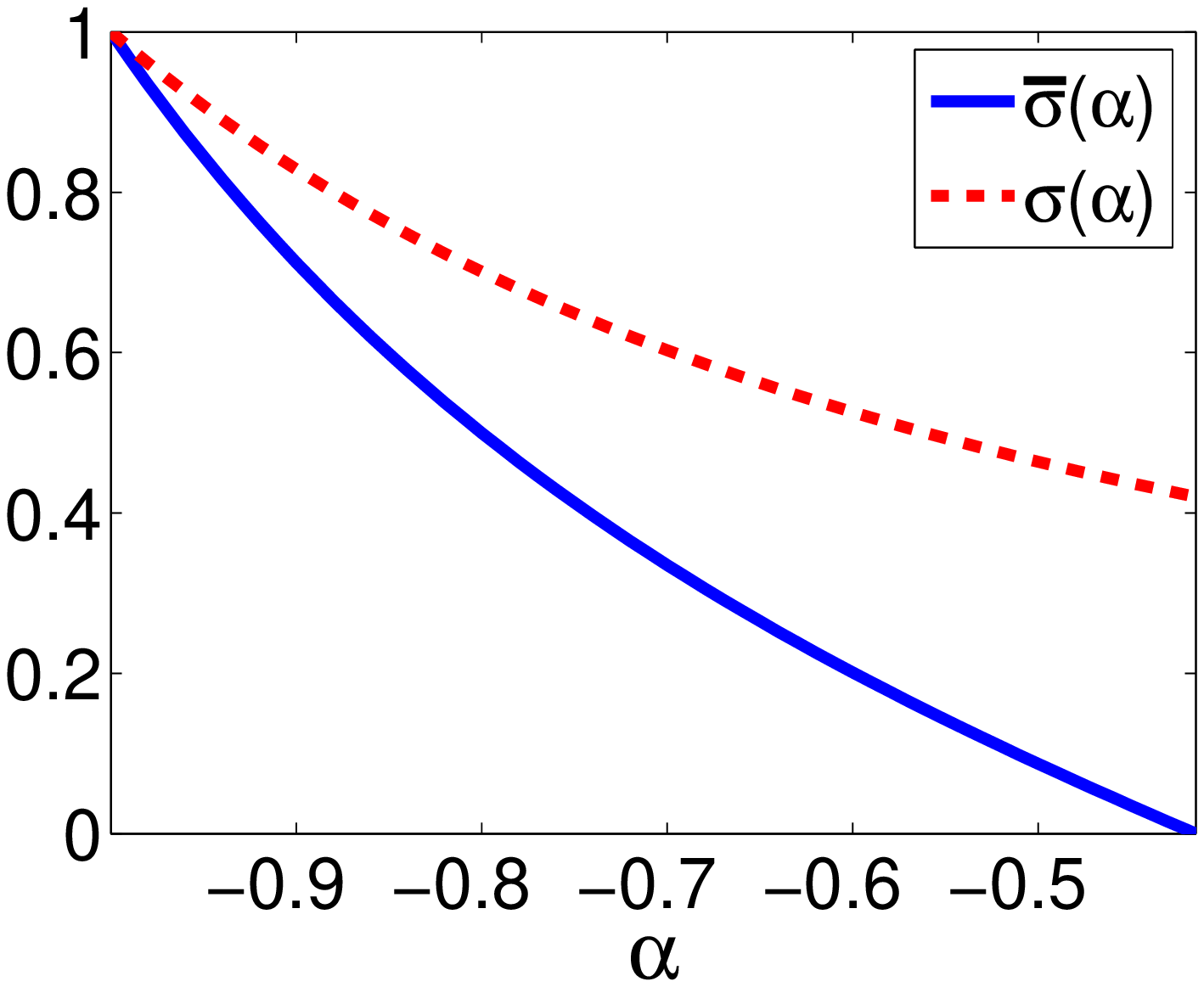}{2.2in}
\caption{(a) {\it
The graph of the exponent $\overline{\omega}(\alpha)$ given
by $(\ref{scalingAtgenRSA})$. The dashed line shows the
exponent $\omega(\alpha)$ given
by $(\ref{scalingAt})$.}
(b) {\it
The graph of the exponent $\sigmabar(\alpha)$ given
by $(\ref{scalingNtgenRSA})$. The dashed line shows the
exponent $\sigma(\alpha)$ given
by $(\ref{scalingNt})$.}
}
\label{figgraphomegasigmabar}
\end{figure}

Next, we will study the behaviour of the system for 
$\alpha < \overline{\alpha}$.
Here, we will assume the scaling ansatz (\ref{scalingansatz}).
Differentiating (\ref{defHgammaalpha})
with respect of $\gamma$, we obtain
\begin{equation}
\frac{\partial H}{\partial \gamma}
(\gamma,\alpha)
=
\frac{2^{\alpha+2}-4}{\alpha} 
B(\gamma+1,\alpha+2) 
\Big[ \psi_0 (\gamma+1) - \psi_0 (\gamma+\alpha+3) \Big]
\label{derHgammaalpha}
\end{equation}
where $\psi_0$ is the polygamma function. For each 
$\alpha < \overline{\alpha}$, the equation
\begin{equation}
H(\gammabar,\alpha) = 0
\label{equationHzero}
\end{equation}
defines implicitly
the function $\gammabar(\alpha)$. 
If $\gamma$ is equal
to $\gammabar(\alpha)$, then the moment $M_\gamma$
is independent of time. Hence, using (\ref{scalmoments}),
we obtain the relation 
$$
a = b + b \, \gammabar(\alpha)
$$
between the coefficients of the scaling ansatz (\ref{scalingansatz}) and 
the parameter $\alpha$ of the model.
Finally, substituting the scaling ansatz (\ref{scalingansatz})
into (\ref{evolmom1}), one can find that
$b = (\alpha+2)^{-1}.$ 
Hence, the scaling of moments
(\ref{scalmoments}) can be rewritten in the form 
\begin{equation}
M_\beta (t) \sim  
t^{\mubar}
\qquad
\mbox{where}
\quad
\mubar
=
\frac{\gammabar(\alpha) - \beta}{\alpha+2}. 
\label{scalmomentsrevisedgenRSA}
\end{equation}
Using (\ref{evolA}) and (\ref{scalmomentsrevisedgenRSA}), we obtain
\begin{equation}
1 - A(t) 
=
\int_0^1 x G(x,t) \dx 
\sim
t^{-\omegabar(\alpha)}
\qquad
\mbox{where}
\quad
\omegabar(\alpha)
=
\frac{1-\gammabar(\alpha)}{\alpha+2}. 
\label{scalingAtgenRSA}
\end{equation}
The graph of the function $\omegabar(\alpha)$
is given in Figure \ref{figgraphomegasigmabar}(a).
We also plot $\omega(\alpha)$ given by (\ref{scalingAt})
for comparison.
Using (\ref{equationforN2}) and (\ref{scalmomentsrevisedgenRSA}),
we also obtain
\begin{equation}
N(t)
\sim  
t^{\sigmabar(\alpha)}
\qquad
\mbox{where}
\;
\sigmabar(\alpha)
=
\frac{\gammabar(\alpha)}{\alpha+2}. 
\label{scalingNtgenRSA}
\end{equation}
The graph of the function $\sigmabar(\alpha)$ is given in Figure 
\ref{figgraphomegasigmabar}(b); we also plot 
$\sigma(\alpha)$ given by (\ref{scalingNt})
for comparison.

\begin{figure}
\centerline{{\large $\alpha = - \displaystyle\frac{1}{2}$}
\qquad \qquad \qquad \qquad \qquad \qquad \qquad 
{\large $\alpha = - \displaystyle\frac{2}{3}$}}
\centerline{
\psfig{file=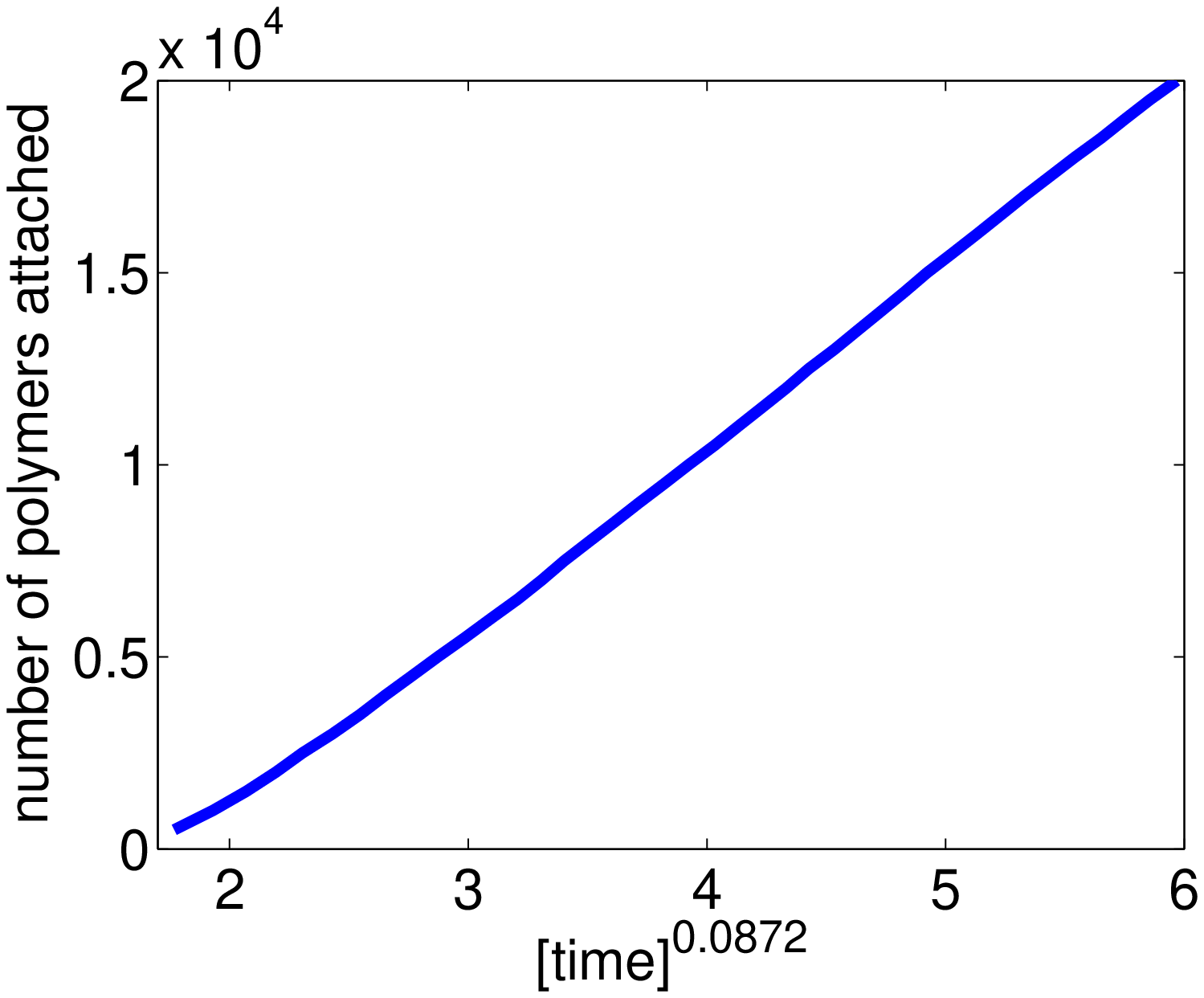,height=2.2in}
\quad
\psfig{file=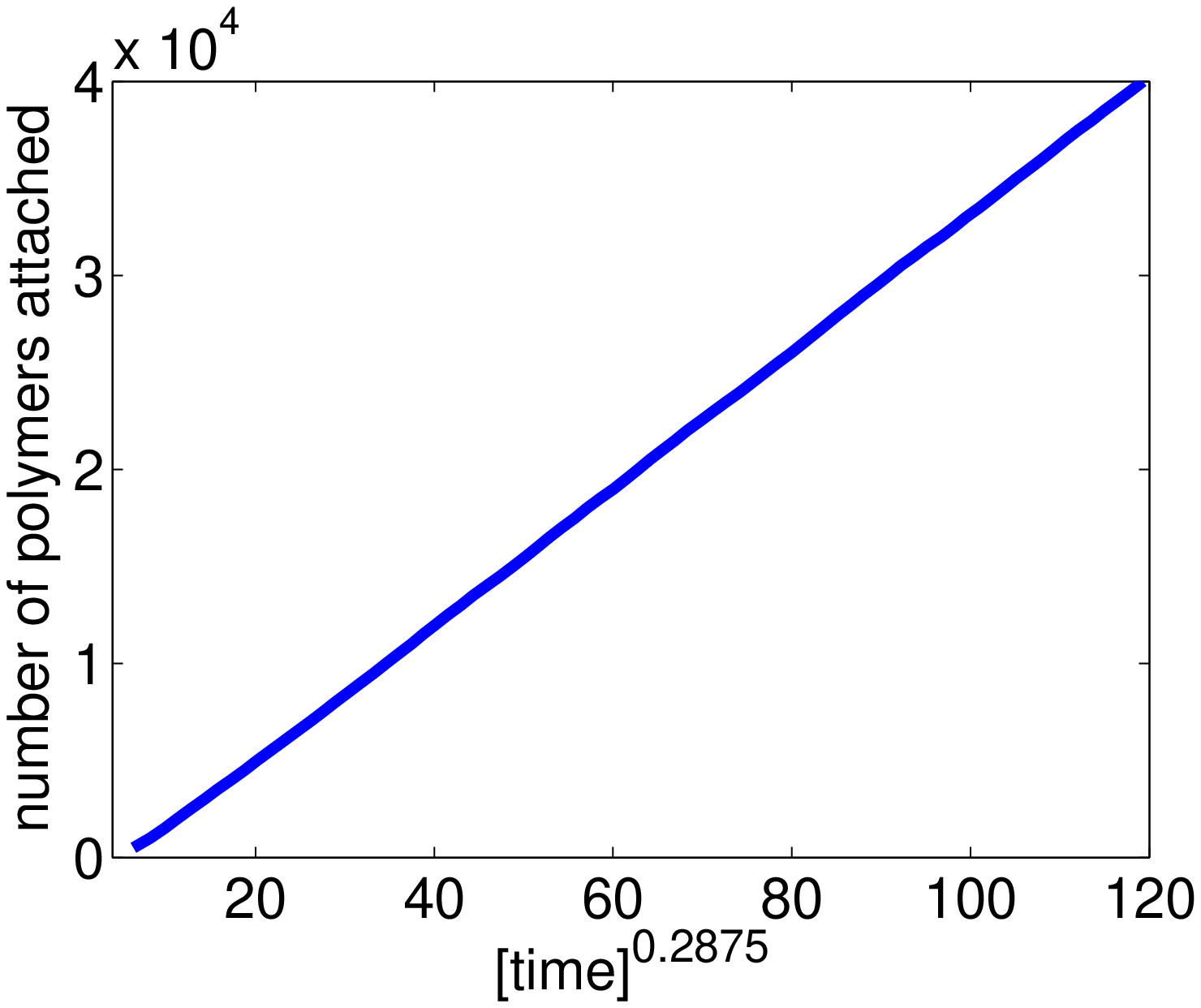,height=2.2in}
}
\centerline{
\psfig{file=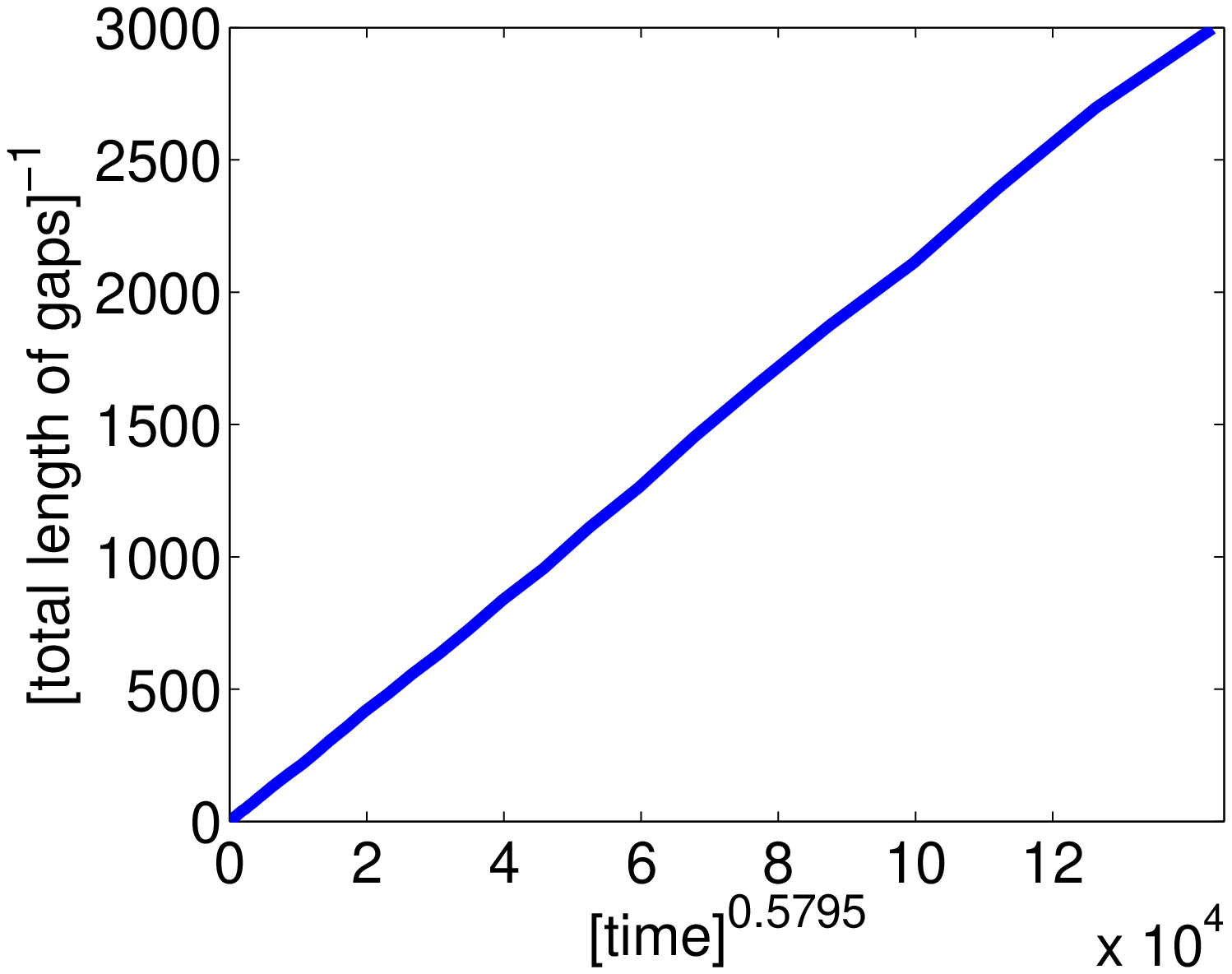,height=2.2in}
\quad
\psfig{file=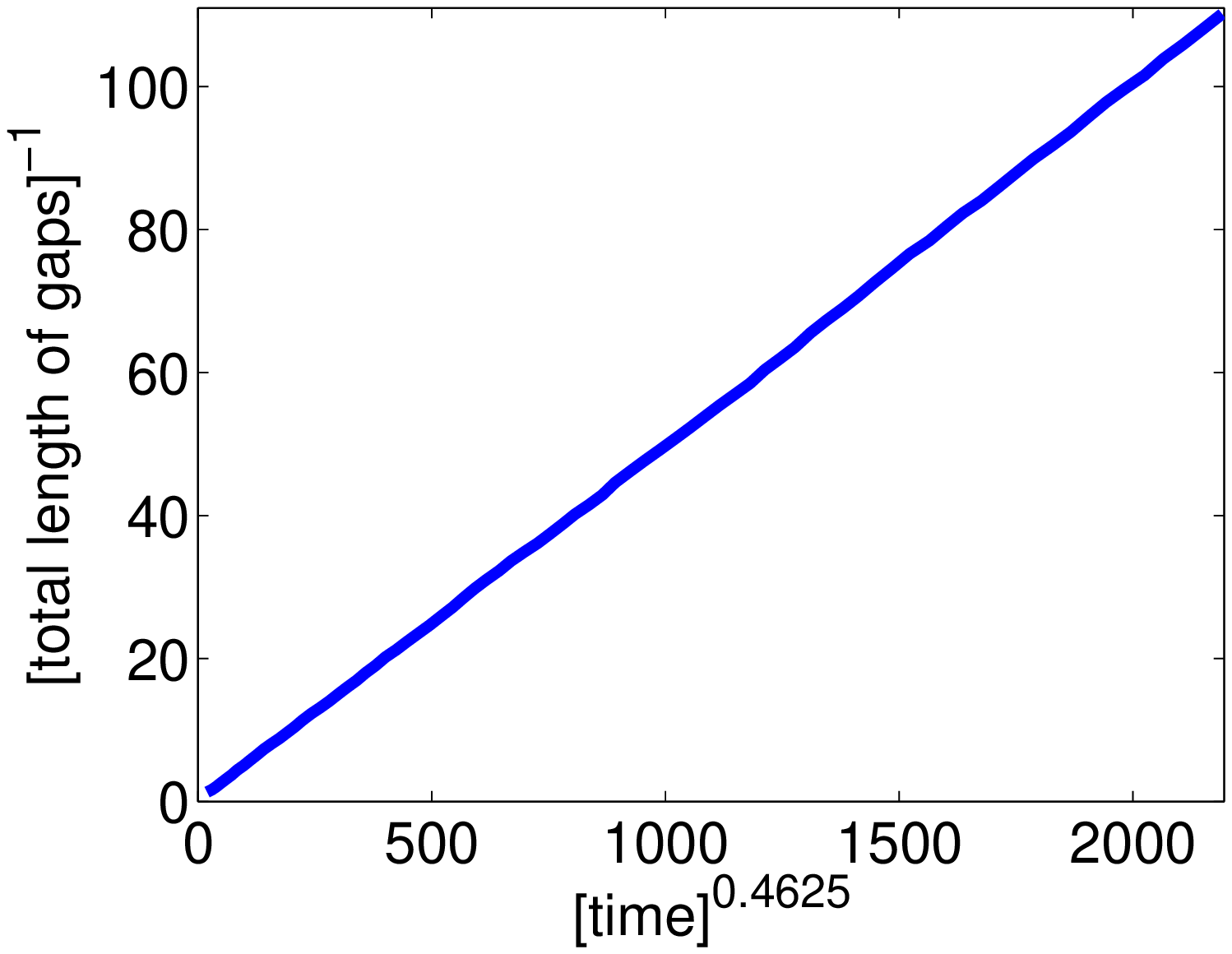,height=2.2in}
}
\centerline{
\psfig{file=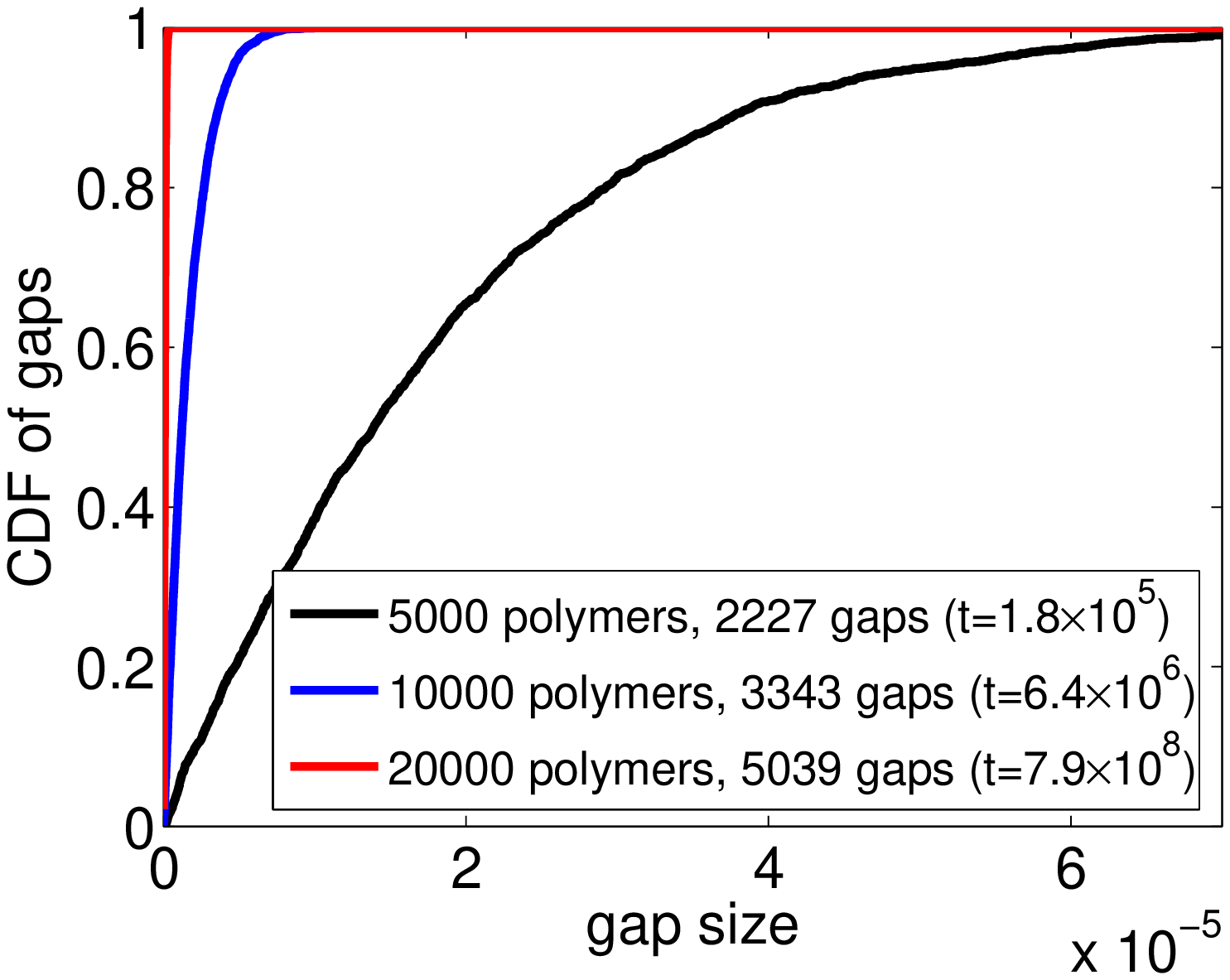,height=2.2in}
\quad
\psfig{file=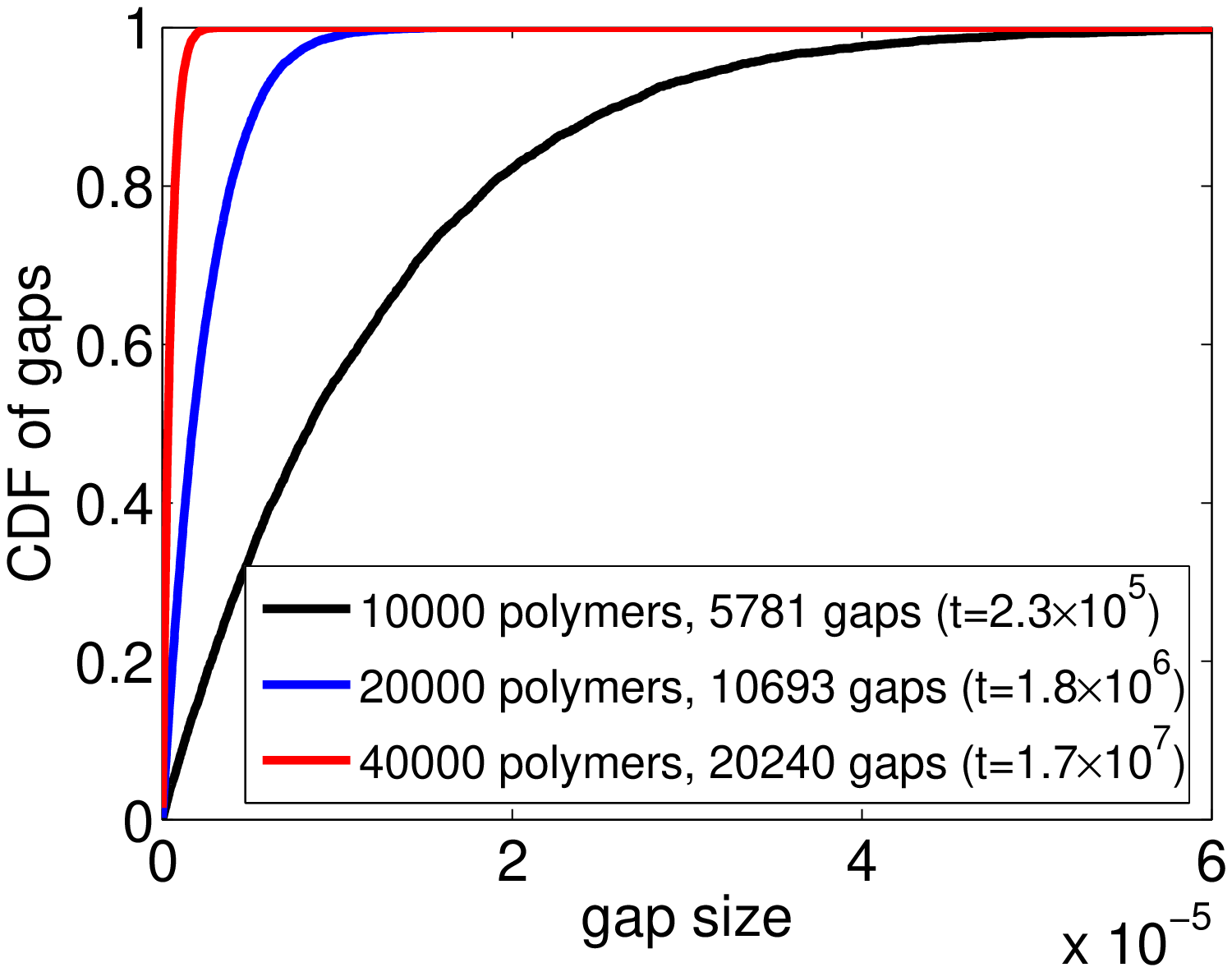,height=2.2in}
}
\caption{{\it gRSA model for $\alpha=-0.5$ (panels on
the left) and $\alpha=-2/3$ (panels on the right).
We plot the time evolution of the number of adsorbed
polymers (top panels) and the time evolution of the
inverse gap size 
$[1-A(t)]^{-1}$ (middle panels). Time is scaled according
to $(\ref{scalingexample})$ (top and middle panels).
We also plot the cumulative distribution function
$C(x,t)$ at different times for both simulations
(bottom panels). 
}}
\label{figalphalarge1}
\end{figure}

To illustrate the results (\ref{scalingAtgenRSA}) and
(\ref{scalingNtgenRSA}), we will execute two gRSA
stochastic simulations. 
We will use
(\ref{distrlength}), (\ref{defxi1}) and (\ref{defxi2})
where $\alpha=-0.5$ or $\alpha=-2/3$. We choose
$\varepsilon=10^{-3}$. 
The results are given
in Figure \ref{figalphalarge1} where
the time evolution of the number of gaps and
the number of adsorbed polymers are shown.
The time is scaled according to (\ref{scalingAtgenRSA}) and
(\ref{scalingNtgenRSA}); we solve
(\ref{equationHzero}) to obtain the desired exponents
\begin{equation}
\sigmabar(-0.5) =0.0872,
\quad 
\omegabar(-0.5) =0.5795,
\quad 
\sigmabar\left(- \frac{2}{3}\right) =0.2875,
\quad 
\omegabar\left(- \frac{2}{3}\right) =0.4625,
\label{scalingexample}
\end{equation}
and then we plot the results of stochastic simulations using the
corresponding scaling (\ref{scalingexample}).
\begin{figure}
\centerline{{\large $\alpha = - \displaystyle\frac{1}{2}$}
\qquad \qquad \qquad \qquad \qquad \qquad \qquad 
{\large $\alpha = - \displaystyle \frac{2}{3}$}}
\centerline{
\psfig{file=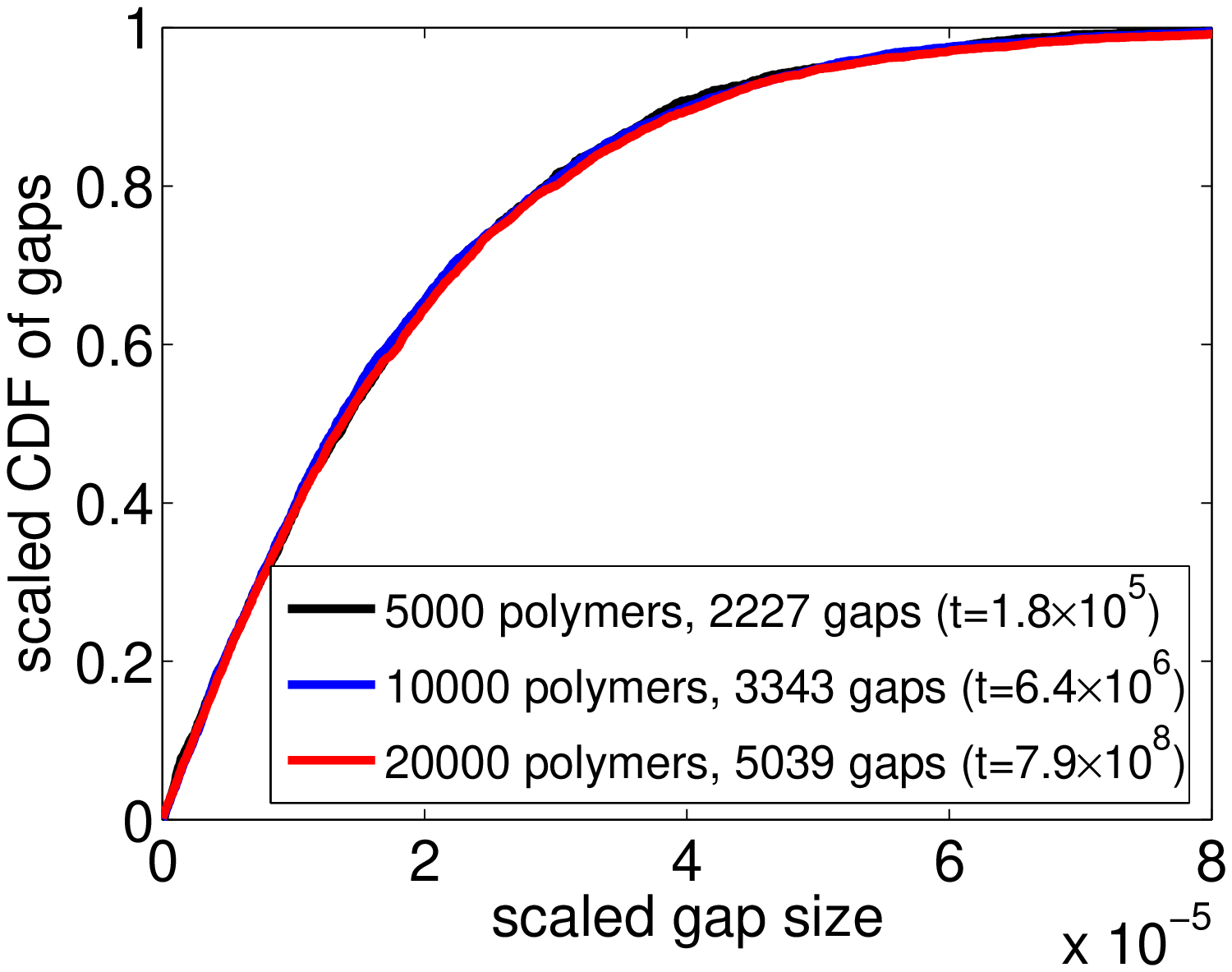,height=2.2in}
\quad
\psfig{file=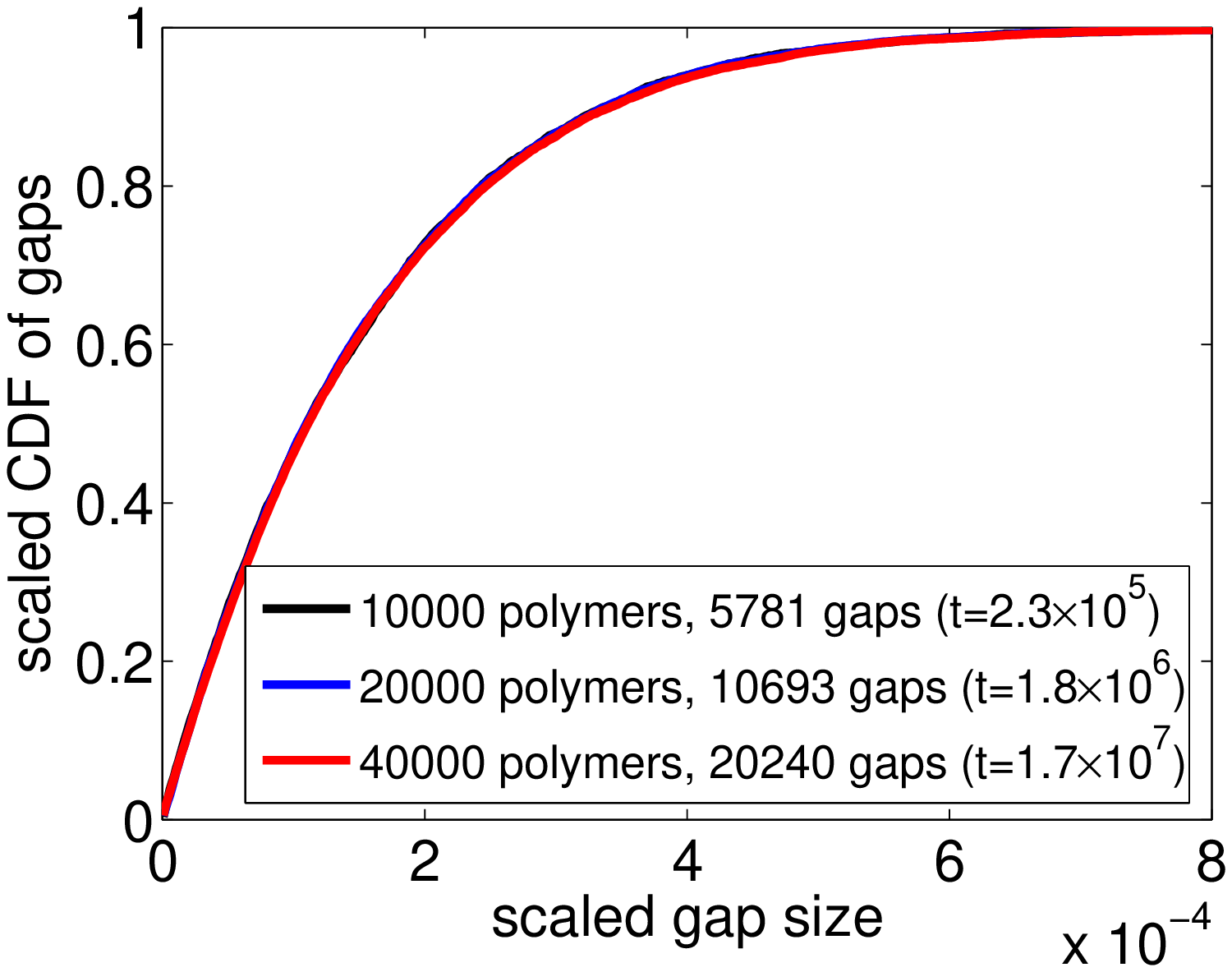,height=2.2in}
}
\caption{{\it Scaled cumulative
distribution function of gaps for gRSA model for $\alpha=-0.5$ 
(panel on the left) and $\alpha=-2/3$ (panel on the right).
}}
\label{figalphalarge2}
\end{figure}
In Figure \ref{figalphalarge1}, we also 
plot the cumulative distribution function $C(x,t)$
for different times (i.e. for different numbers of polymers
attached). 
Using the suitable rescaling $\widehat{C}(x,t) = C(kx,t)$, 
the curves collapse to a single curve as shown in
Figure \ref{figalphalarge2}.

Finally, one can easily show that formula (\ref{scalingAtgenRSA}) 
works also for the case $\overline{\alpha} \le \alpha < 0.$ 
To illustrate it, we plot the time evolution of the quantity 
$[1-A(t)]^{-1}$ for $\alpha=-0.1$ and $\alpha=-0.3$
in Figure \ref{figalphasmall} (bottom panel) using the scaling
(\ref{scalingAtgenRSA}). On the other hand, formula
(\ref{scalingNtgenRSA}) is no longer correct because
integrating of (\ref{equationforN2}) implies that 
$N(t)$ is approaching a constant value. More precisely,
$N(t) \sim C + t^{\sigmabar(\alpha)}$ for
$\sigmabar(\alpha)<0.$ If $\alpha$ is positive than
(\ref{decaygaps}) is valid, i.e. we have (\ref{decaygaps})
for any $\alpha>\overline{\alpha}$.

\subsection{Time dependent concentration of reactive polymers}

\label{sectimedependent}

As discussed before, the reactive groups on our polymers
are capable of reacting with the solvent before reaching the 
surface \cite{Subr:2005:CDC}. 
It may therefore be more realistic 
to consider that only a fraction of polymers
$r(t)$ is still reactive at time $t$. 
Here, $r(t) \in [0,1]$,
$r(0)=1$ and $r(t)$ is a decreasing function of time.

The random sequential adsorption algorithm has to be modified as
follows: at each time step, we generate the random number
uniformly distributed in the interval $(0,1)$.
If this number is greater than $r(t)$, then the selected
polymer  has lost its binding site through reaction
with the solvent (it cannot
be adsorbed), and we continue with the next step. 
Otherwise,
we choose randomly a position on the interval and we attempt
to place the polymer there.

\begin{figure}[t]
\centerline{
\psfig{file=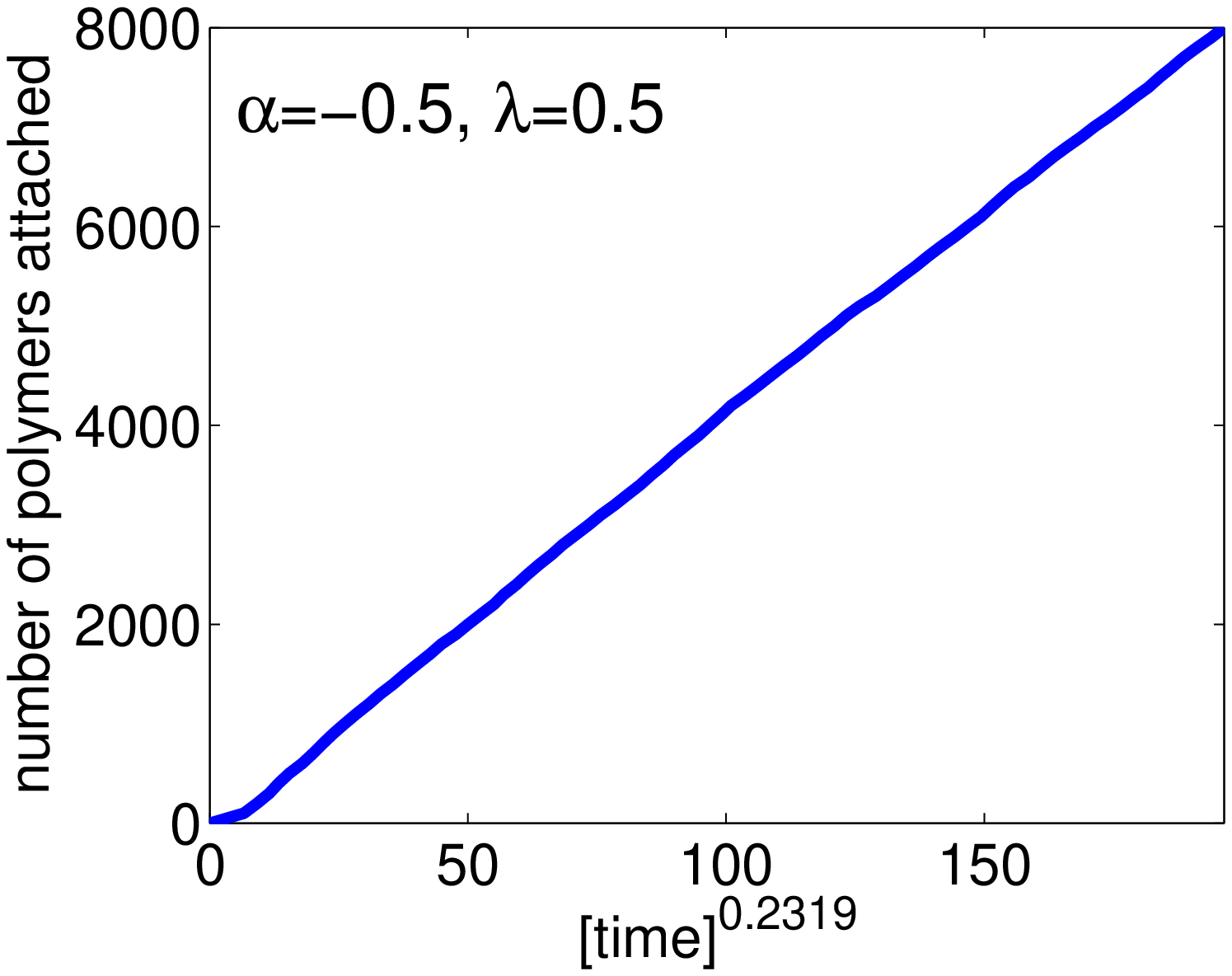,height=2.1in}
\qquad
\psfig{file=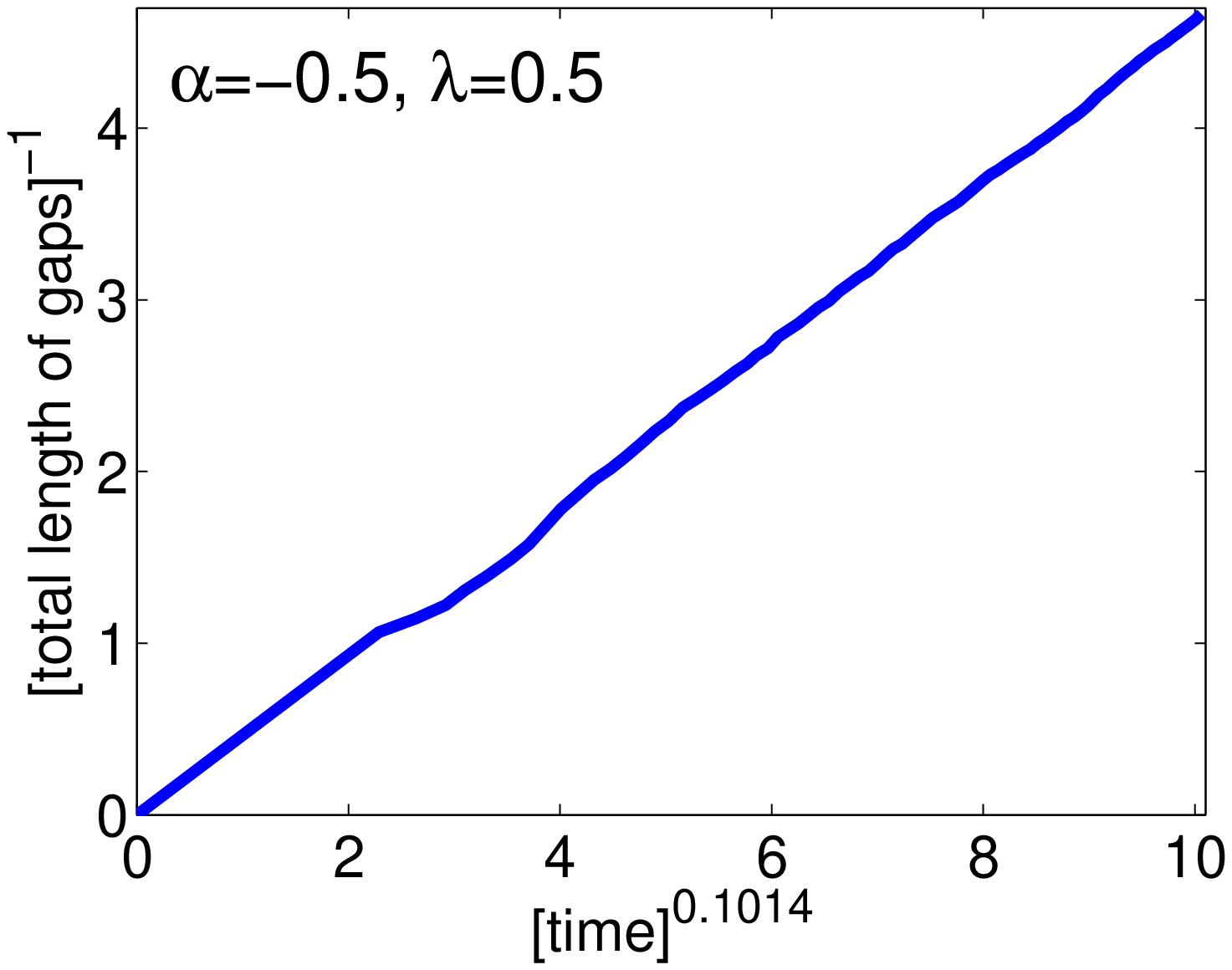,height=2.1in}
}
\centerline{
\psfig{file=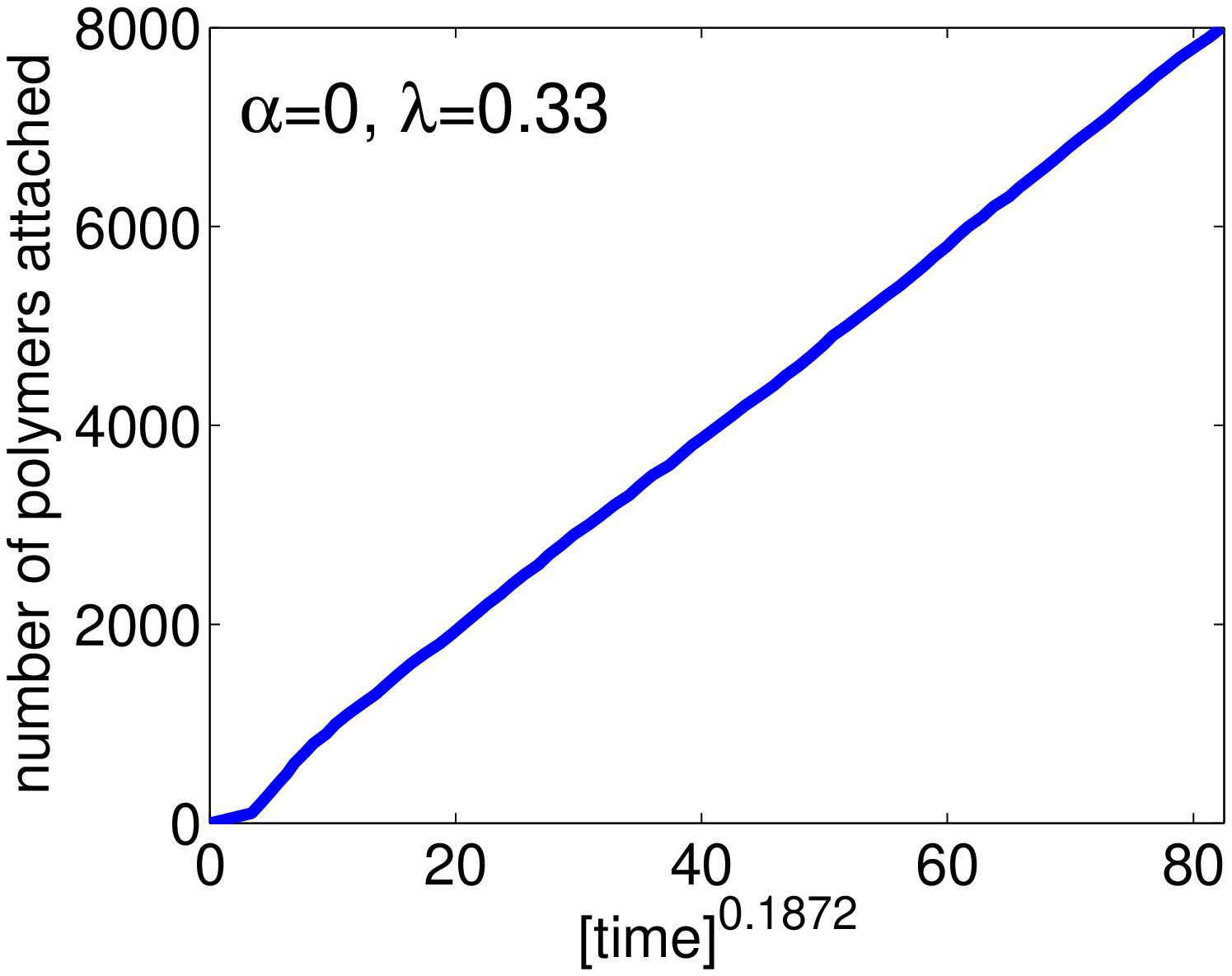,height=2.1in}
\qquad
\psfig{file=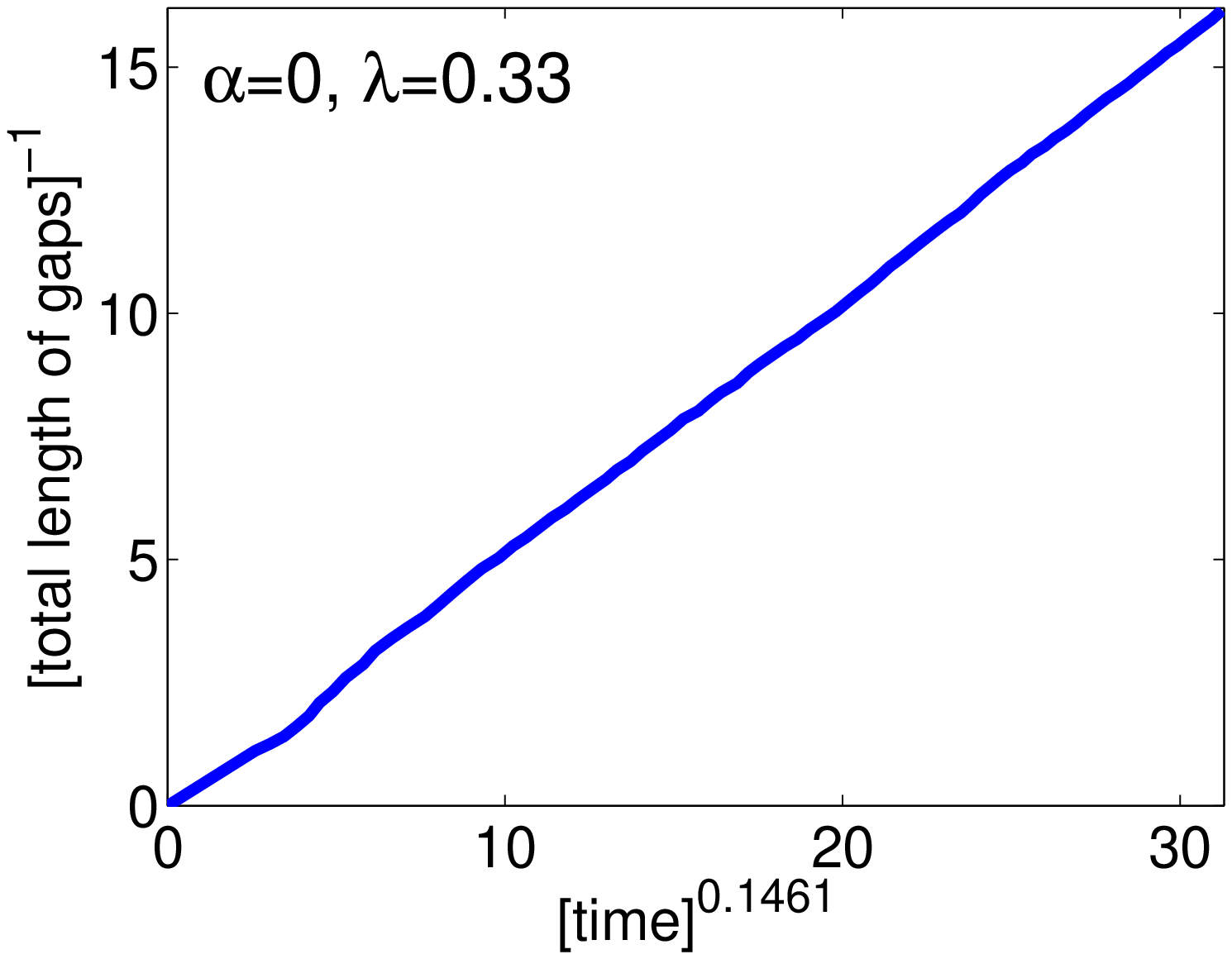,height=2.1in}
}
\caption{{\it Modified RSA model from Section $\ref{sectimedependent}.$
The time evolution of the number of polymer molecules attached
to the surface $N(t)$} (left). {\it Time evolution of the inverse of the
total gap size $[1-A(t)]^{-1}$} (right). {\it Time is scaled according
to $(\ref{scalingANtlambda})$.
}}
\label{figpolattgap2}
\end{figure}

Depending on the form of function $r(t)$, different dynamics
can be observed. 
First, let us suppose that 
\begin{equation}
r(t) = \frac{1}{t^{\lambda}}
\qquad
\mbox{for} \; \lambda \in [0,1).
\label{formr1}
\end{equation}
In this case, we can find a relation between the modified random
sequential algorithm and the previous results. 
At each time $t$, we can compute the average waiting
time $\Delta t$ before a reactive polymer 
hits the surface as the solution of the
equation
\begin{equation}
\int_{t}^{t+\Delta t} \frac{1}{\tau^{\lambda}} \; \dtau = 1.
\label{equationforDeltat}
\end{equation}
Solving (\ref{equationforDeltat}), we find
$$
(t+\Delta t)^{1-\lambda} = t^{1-\lambda} + 1 - \lambda.
$$
Hence,
$$
\Delta t = \Big[t^{1-\lambda} + 1 - \lambda \Big]^{1/(1-\lambda)} - \, t
\quad
\sim 
\quad
t^{\lambda}. 
$$
Consequently, we can make use of the formulas (\ref{scalingAt})
and (\ref{scalingNt}), or formulas (\ref{scalingAtgenRSA})
and (\ref{scalingNtgenRSA}), in the case (\ref{formr1}).
For example, using
(\ref{scalingAt}) and (\ref{scalingNt}), we obtain
that the quantities $A(t)$ and $N(t)$ 
satisfy the following asymptotic behaviour
\begin{equation}
A(t)
\sim  
t^{(1-\lambda)\omega(\alpha)}
\qquad
\mbox{and}
\qquad
N(t)
\sim  
t^{(1-\lambda)\sigma(\alpha)}.
\label{scalingANtlambda}
\end{equation}
To illustrate the formula (\ref{scalingANtlambda}),
we stochastically simulate the 
gRSA model with the probability distribution $p(z)$ given 
by (\ref{distrlength}) and the probability $\xi(z,w-x_1,x_2-w)$ given 
by (\ref{defxi3}), where  the fraction of the reactive polymers 
in the system decreases with time according 
to (\ref{formr1}). 
We select $\varepsilon=10^{-3}$ and verify the asymptotic behaviour
(\ref{scalingANtlambda}) for
$\alpha=-0.5$ and $\lambda=0.5$. 
Then (\ref{scalingANtlambda}) implies
$$
1 - A(t) \sim t^{-0.1014}
\qquad
\mbox{and}
\qquad
N(t) \sim t^{0.2319}.
$$ 
The time evolution
of $A(t)$ and $N(t)$ is given in Figure
\ref{figpolattgap2} (top panels).
We also present results for
$\alpha = 0$ and $\lambda=0.33$
in Figure \ref{figpolattgap2} (bottom panels).
Again, we scale the time according to 
(\ref{scalingANtlambda}). 

The reactive group is lost by chemical reaction with the solvent. 
It might be more natural to consider (instead of 
(\ref{formr1})) that the fraction of
reactive polymers is exponentially decreasing, i.e.
\begin{equation}
r(t) = e^{- \lambda t}
\qquad
\mbox{for} \; \lambda > 0.
\label{formr2}
\end{equation}
The formula (\ref{formr2}) gives rise to qualitatively different 
dynamics for the system, as opposed to the dynamics associated with
(\ref{formr1}). 
For simplicity, let us assume that every
polymer with a functional reactive group can be adsorbed
(which will give a bound on $N(t)$ from above).
Then the average number of adsorbed polymers $N(t)$ is
$$
N(t)=\frac{1}{\lambda} \Big(1 - e^{- \lambda t} \Big),
$$
which implies that the number of adsorbed polymers does not approach
infinity as in the previous case. 

\bigskip

\section{Equation-free analysis of gRSA}

\label{seceqfree}

\begin{figure}
\centerline{\psfig{file=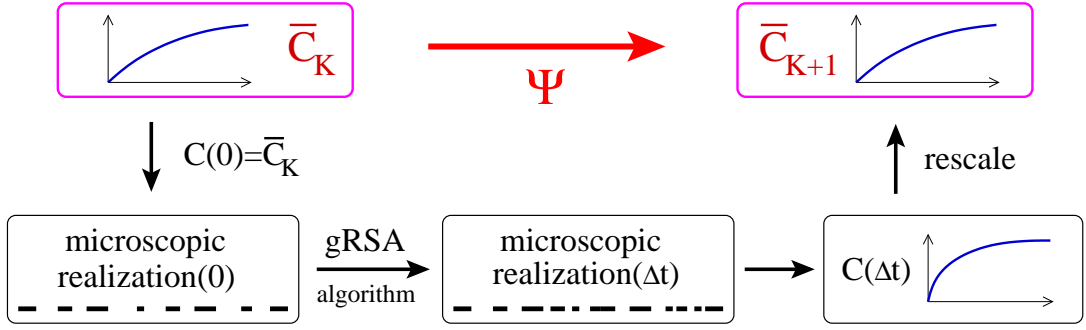,height=1.7in}}
\caption{{\it
Schematic of the equation-free mapping $\Psi$.}
}
\label{figeqfreescheme}
\end{figure}

In the previous theory, we assumed the scaling ansatz (\ref{scalingansatz})
(see also \cite{Krapivsky:1992:KRS})
for  $G(x,t)$ 
and we computed the time dependence of the quantities of interest
$A(t)$ and $N(t)$. 
A related interesting question is whether we can also compute the profile 
$\Phi$ from (\ref{scalingansatz}). One possibility is to
substitute (\ref{scalingansatz}) in equation (\ref{evolGgen4})
and solve it numerically for $\Phi$ but we will not proceed
this way.
Instead, we demonstrate the computation
of self-similar profile $\Phi$ using only short-time
appropriately initialized simulations of the stochastic gRSA model.
In this equation-free context \cite{Kevrekidis:2003:EFM,Chen:2005:EFD}, 
it is easier to work with the
cumulative distribution function $C(x,t)$, which can be obtained
from $G(x,t)$ through (\ref{forcdf}); $C(x,t)$ is less noisy than $G(x,t)$
(e.g. \cite{Gear:2001:PIM}). 
Using (\ref{forcdf}) and 
(\ref{scalingansatz}), we obtain
$$
C(x,t) 
= 
\frac{1}{\int_0^\infty G(y,t) \dy} \int_0^x G(y,t) \dy
=
$$
$$
= 
\frac{1}{\int_0^\infty \Phi \left( y \, t^b \, \right) \dy} 
\int_0^x \Phi \left( y \, t^b \, \right) \dy
=
\frac{1}{\int_0^\infty \Phi \left( \xi \right) \dxi} 
\int_0^{x t^{b}} \Phi \left( \xi \right) \dxi.
$$
Hence, we see that the cumulative density function
$C(x,t)$ scales as 
\begin{equation}
C(x,t) \equiv  \overline{C}(x t^b).
\label{forcdfscaling}
\end{equation}
To compute the profile $\overline{C}$, we can use
an equation-free iterative fixed point algorithm which is shown
schematically in Figure \ref{figeqfreescheme}.
Starting with the initial guess $\overline{C}_0$,
we compute the sequence of profiles
$\overline{C}_K$, $K=1, 2, 3, \dots$, where
\begin{equation}
\overline{C}_{K+1} = \Psi (\overline{C}_{K}),
\qquad
\mbox{for} \; K=0, 1, 2, 3, \dots,
\label{formulapsi}
\end{equation}
and where the mapping $\Psi$ is obtained as the composition
of the following four steps:

\bigskip

\leftskip 1cm

\noindent
(a) Given the cumulative density profile $\overline{C}_K$, 
create one or more  microscopic realizations of gaps in the unit interval
such that the initial cumulative density function 
is $C(\cdot,0)=\overline{C}_K$.

\smallskip

\noindent
(b) Use the microscopic simulator (i.e. use the gRSA algorithm) 
for a short time $\Delta t.$

\smallskip

\noindent
(c) Compute the new cumulative distribution function
$C(\cdot,\Delta t)$ at time $\Delta t.$

\smallskip

\noindent
(d) Rescale $C(\cdot,\Delta t)$ to compute $\overline{C}_{K+1}$. 

\leftskip 0cm 

\bigskip

\begin{figure}
\centerline{{\large $\alpha = - \displaystyle\frac{1}{2}$}
\qquad \qquad \qquad \qquad \qquad \qquad \qquad 
{\large $\alpha = - \displaystyle \frac{2}{3}$}}
\centerline{
\psfig{file=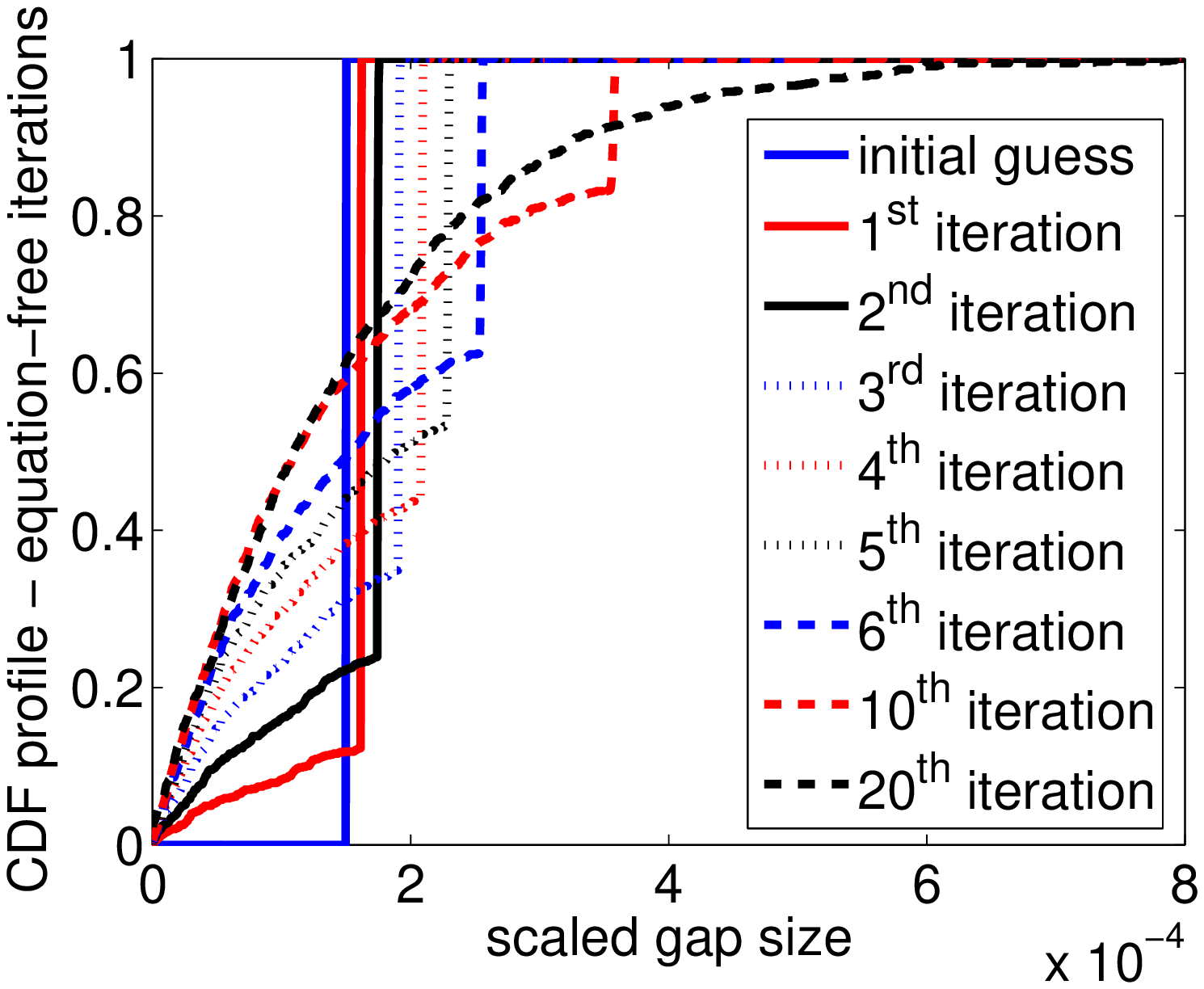,height=2.2in}
\quad
\psfig{file=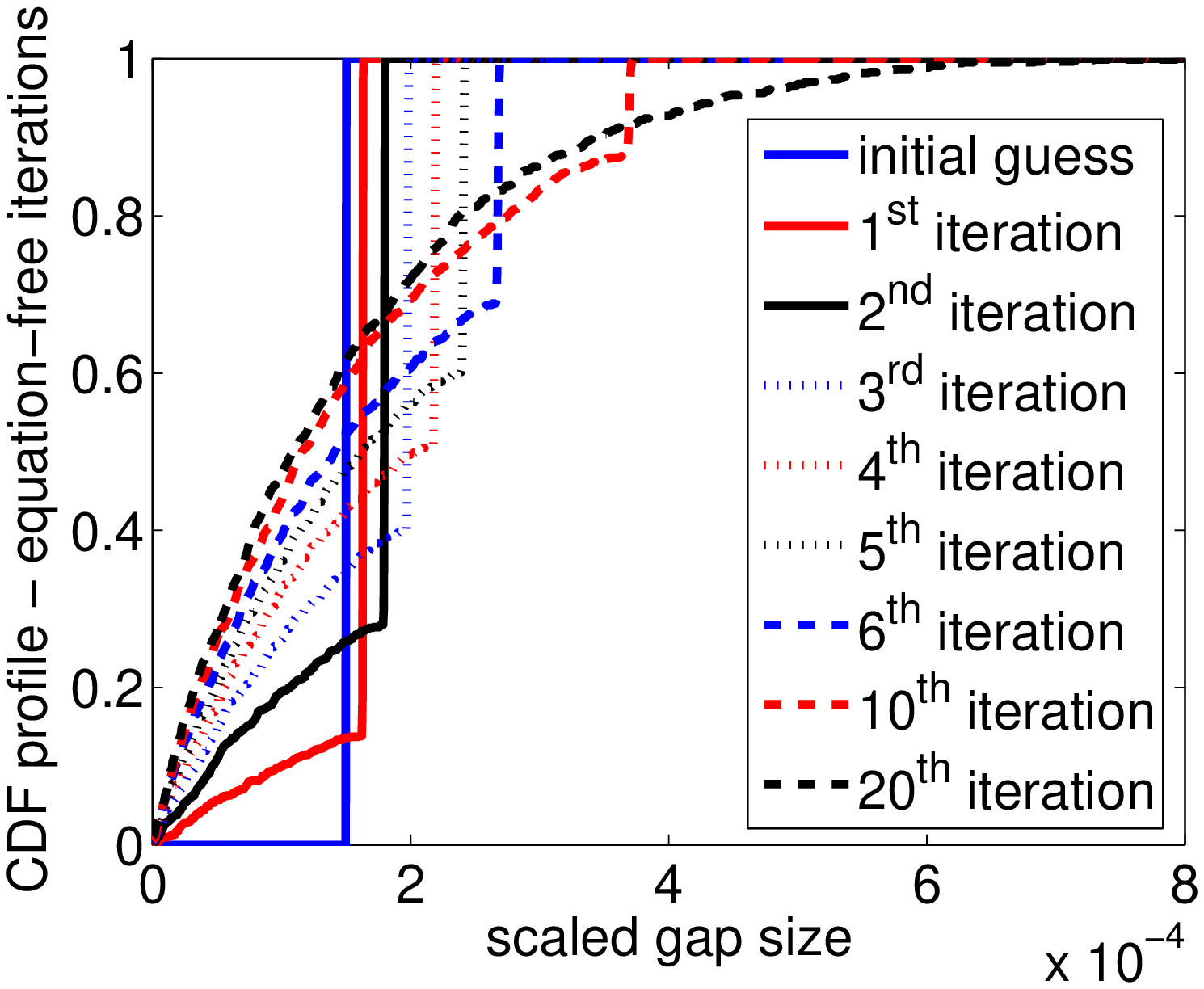,height=2.2in}
}
\smallskip
\centerline{
\psfig{file=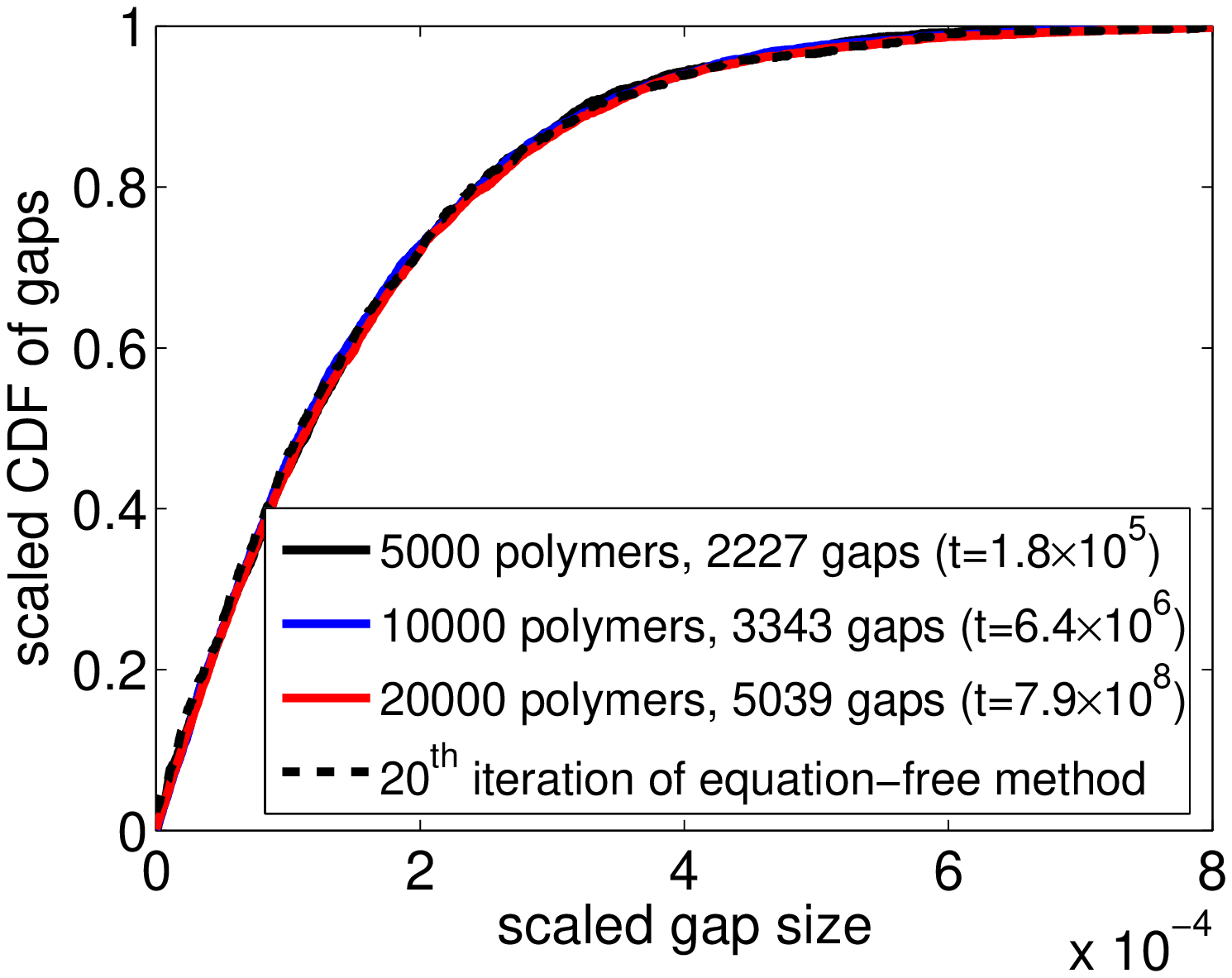,height=2.2in}
\quad
\psfig{file=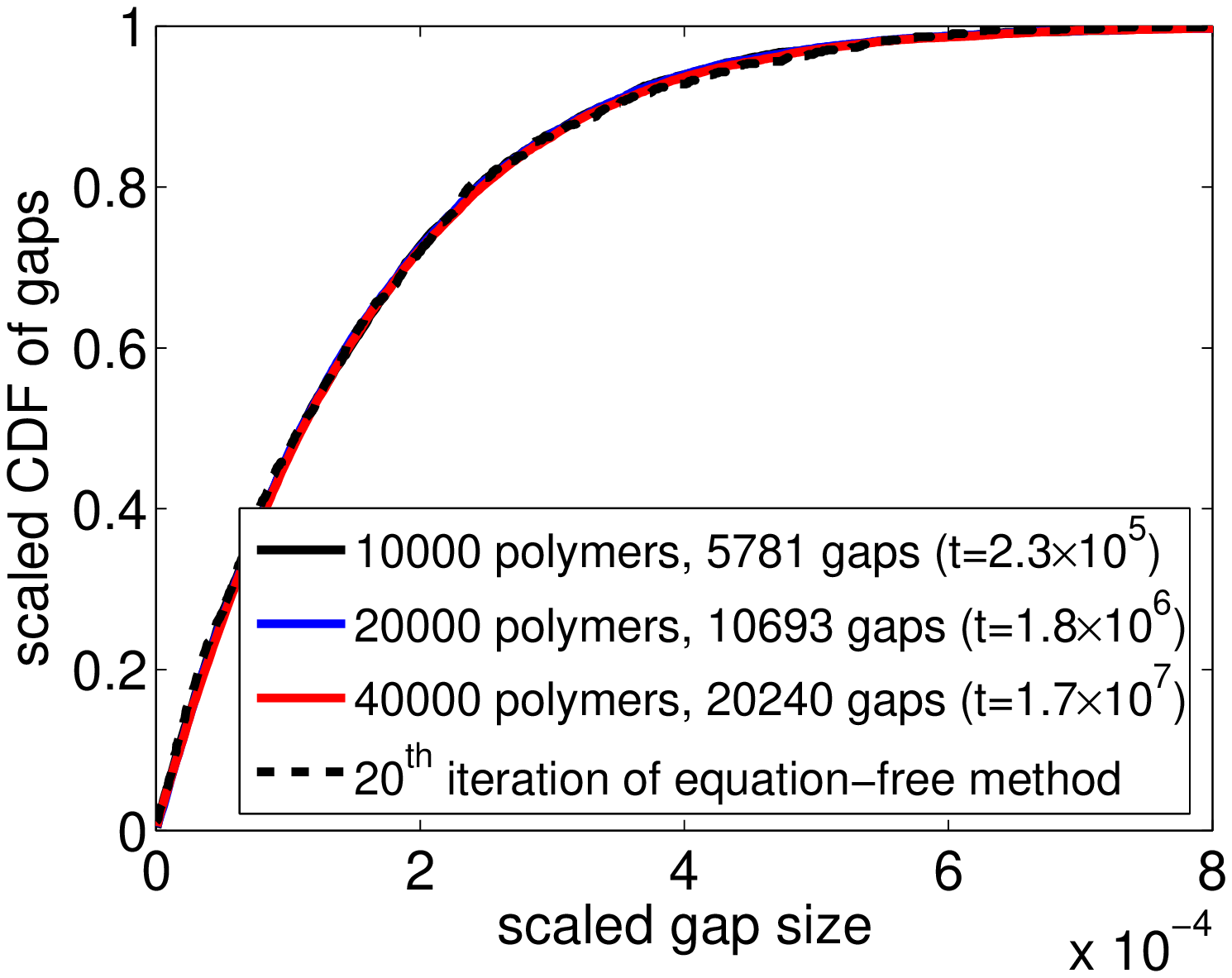,height=2.2in}
}
\caption{{\it Equation free gRSA computational results
for $\alpha=-0.5$ (left panels) and $\alpha=-2/3$
(right panels).
Iterations of the equation-free dynamic renormalization algorithm
$(\ref{formulapsi})$ (top panels). 
Comparison with the steady steate profile obtained through long time
simulations (bottom panels).}
}
\label{figeqfreeresults}
\end{figure}

\noindent
One possible way to rescale $C(\cdot,\Delta t)$ is to compute the 
average gap size $a_0$ from $C(\cdot,0)$ and the
average gap size $a_{\Delta t}$ from $C(\cdot,\Delta t)$.
Then the $\overline{C}_{K+1}$ can be computed by
\begin{equation}
\overline{C}_{K+1}(x) 
= 
C \left( \frac{a_{\Delta t}}{a_0} x ,\Delta t \right).
\label{rescaling}
\end{equation}
We now present illustrative results obtained by
this fixed point computation (\ref{formulapsi}) using the gRSA algorithm. 
We will use
(\ref{distrlength}), (\ref{defxi1}) and (\ref{defxi2}) 
where $\alpha=-0.5$ or $\alpha=-2/3$. 
We choose $\epsilon = 10^{-3}$. 
The results of long
term simulations for these parameter values were already
shown in Figure \ref{figalphalarge1}. 
Our goal is to use
the iterative formula (\ref{formulapsi}) to compute
the scaled cumulative distribution function profile which
was shown in Figure \ref{figalphalarge2}.
This algorithm allows us to find the self-similar shape
by performing simulations while simulating at a scale 
(at relatively larger average gap sizes) where the evolution
is relatively fast, compared to the long-term dynamics
close to jamming.
The initial guess is given as 
$$
\overline{C}_0(x)
=
\left\{
\begin{array}{ll}
0 \quad & \mbox{for} \; x \le 1.5 \times 10^{-4}; \\
1 \quad & \mbox{for} \; x > 1.5 \times 10^{-4}; \\ 
\end{array}
\right.
$$
which means that initially all our gaps have the same size
$1.5 \times 10^{-4}$. 
At each iteration step (see
Figure \ref{figeqfreescheme}), we place 1000 gaps 
according to the cumulative distribution function
$\overline{C}_K$ to the interval $[0,1].$ 
We evolve the
simulation until 100 new polymers are placed. 
We then {\it rescale} the new cumulative distribution according
to (\ref{rescaling}) and we compute $\overline{C}_{K+1}$.
Several first iterations are shown in Figure \ref{figeqfreeresults} 
(top panels). 
We see that after 20 iterations, we have effectively 
reached the steady state (the stationary shape of the self-similarly
evolving gap distribution). 
More precisely, the error between 
iterations is small and it is not further systematically 
decreasing. 
The comparison of the equation-free 20th iteration
with the results obtained by the long-time simulations
are also shown in Figure \ref{figeqfreeresults} (bottom panels).

Finally, we note that many other algorithmic possibilities
for the computation of the profile $\overline{C}$ exist. 
The equation
(\ref{formulapsi}) seeks a fixed point 
of the mapping $\Psi$. 
Instead of successive substitution, other fixed point algorithms
implemented in a matrix-free fashion through short simulation bursts
can be used to find stationary solutions - for example, 
Newton-GMRES iterations \cite{Chen:2005:EFD}; these would be
able to converge on even dynamically unstable self-similarly
evolving distributions.

\section{Discussion}

\label{secdiscussion}

In this paper, motivated by a pharmacological example involving 
polymer coating of a virus surface, 
we studied certain aspects of polydisperse adsorption
of macromolecules in one spatial dimension.
We presented an extension of the
classical random sequential adsorption algorithm
to capture better certain essential properties of the
pharmacological model system currently used
in drug development research. 
We introduced partial
overlapping of adsorbing macromolecules, i.e. 
we considered that the polymers are not
rigid objects but they can be deformed while
attaching to the surface.  
We found two distinct asymptotic regimes.
Depending on the 
parameters of the processes involved, we can observe that either
(a) the number of gaps between polymers asymptotically approaches zero, or 
that 
(b) the number of gaps asymptotes to infinity and the gap 
distribution acquires an asymptotically self-similar profile.  

We also briefly discussed the impact of a possible reaction 
of the polymers with the
solvent on gRSA dynamics. 
Again, two possibilities exist.
If the decay of the reactive groups is relatively weak, then
the dynamics of the system remains qualitatively unchanged and the
system only evolves on a slower time scale. 
On the other hand,
if the reactive groups decay exponentially, this decay
ultimately wins over the polynomial time asymptotics
of gRSA. 
From the applications point of view, it therefore becomes crucial
to know the corresponding rate constants in order to reliably predict
what type
of behaviour one might expect over the time scales of interest.
Typically, the coating process is performed overnight in the laboratory
and different reactive groups have different half lives; measuring
these rates becomes an important task.

In this paper we worked in one spatial dimension and provided analytical 
results  about the long time behaviour of gRSA models.
The analytical approach was based on 
two important facts: we knew what the good macroscopic observables
for describing the system behaviour were and we were able to write down
analytically tractable equations for these observables.
The good observable for our system was a distribution
of gaps $G(x,t)$ between adsorbed polymers. 
If we know the
initial distribution of gaps $G(0,t)$ one could easily 
predict $G(x,t)$ at future times. 

On the other hand, if we know (hope) that the gap distribution
$G(x,t)$ is a good observable for the system of interest
but we do not know the evolution equation for $G(x,t)$, then it is 
still possible to use the equation-free 
methods \cite{Kevrekidis:2003:EFM}.
The main idea of the equation-free methods is to use the short bursts 
of appropriately initialized microscopic/stochastic 
computations to {\it estimate} macroscopic
quantities of interest on demand. 
Hence, if one does not have an explicit coarse-grained evolution 
equation for the system statistics, one can in principle
avoid long, brute-force simulations.
This might be the case for one-dimensional adsorption
problems with more complicated microscopic evolution rules.

The situation becomes significantly more difficult in the higher dimensional
case. 
Here, the analytical theory is far behind in development,
and the literature contains mostly computational results.
The first question for higher dimensional adsorption is the
nature of the ``good" coarse-grained observables for the system. 
Good 
observables (the variables
in terms of which the unavailable effective model would be written)
are necessary for developing a useful analytical
theory. 
Knowing appropriate coarse-grained observables 
is also an important feature of equation-free algorithms.
Having one-dimensional analogues in mind, we see that one needs
an effective way to describe the statistics of ``gaps" (free space) in higher
dimensions. 
If we cannot estimate (by intuition or by
suitable algorithms for the detection of
low-dimensionality in high-dimensional data) effectively
good observables for the system, then the direct, brute-force computationally
intensive simulations might be the only modelling option. 
In this paper
we showed cases where we could do more than brute-force simulation
and provided 
analytical results giving insights into the dynamics of gRSA.

The problems studied in this paper were motivated 
by the pharmacological example mentioned above, 
and realistic predictive modelling of the problem
clearly requires extensive model parameter information that must
come from experimental data. 
As we showed, we can
expect different dynamics of the problem depending 
on the values of the parameters of the polymer and the virus 
which are used. 
Obtaining reliably such parameters and bounds on their uncertainty
for our particular model problem is non-trivial,
and we are not yet ready to report about it.

\bibliographystyle{amsplain}
\bibliography{bibrad}

\end{document}